\documentclass[graybox]{svmult}

\usepackage{mathptmx}       
\usepackage{helvet}         
\usepackage{courier}        
\usepackage{type1cm}        
\usepackage{makeidx}         
\usepackage{graphicx}        
\usepackage{multicol}        
\usepackage[bottom]{footmisc}

\usepackage{amsmath,comment,amssymb,cite,mathtools}

\newcommand{\R}{\ensuremath{{\mathbb R}}}

\newcommand{\KK}{{\mathcal K}}

\newcommand{\RR}{{\mathcal R}}

\newcommand{\TT}{{\mathcal T}}

\newcommand{\XX}{{\mathcal X}}

\newcommand{\ZZ}{{\mathcal Z}}

\newcommand{\ddd}{{\mathsf d}}
\newcommand{\nn}{{\mathsf{n}}}
\newcommand{\sss}{{\mathsf{s}}}
\newcommand{\uu}{{\mathsf{u}}}

\newcommand{\attack}{{\mathsf{a}}}
\newcommand{\free}{{\mathsf{af}}}
\newcommand{\detect}{{\mathsf{detect}}}

\newcommand{\hazard}{{\mathsf{hazard}}}
\newcommand{\received}{{\mathsf{c}}}

\newcommand{\sat}{\ensuremath{{\rm sat}}}

\newcommand{\ourA}{{O^1}}
\newcommand{\ourB}{{0_1}}

\DeclareMathOperator*{\minimize}{minimize}

\makeindex             

\begin{document}


\title*{Zero-dynamics Attack, Variations, and Countermeasures}
\author{Hyungbo Shim, Juhoon Back, Yongsoon Eun, Gyunghoon Park, and Jihan Kim}
\institute{Hyungbo Shim \at Seoul National University, Seoul, Korea, \email{hshim@snu.ac.kr}
\and Juhoon Back \at Kwangwoon University, Seoul, Korea, \email{backhoon@kw.ac.kr}
\and Yongsoon Eun \at DGIST, Daegu, Korea, \email{yeun@dgist.ac.kr}
\and Gyunghoon Park \at KIST, Seoul Korea, \email{gyunghoon.p@gmail.com}
\and Jihan Kim \at Seoul National University, Seoul, Korea, \email{jihan89@cdsl.kr}
}
%
%
\maketitle


\abstract{This chapter presents an overview on actuator attacks that exploit zero dynamics, and countermeasures against them. First, zero-dynamics attack is re-introduced based on a canonical representation called normal form. Then it is shown that the target dynamic system is at elevated risk if the associated zero dynamics is unstable. From there on, several questions are raised in series  to ensure  when the target system is immune to the attack of this kind. The first question is: {\em Is the target system secure from zero-dynamics attack if it does not have any unstable zeros?} An answer provided for this question is: {\em No, the target system may still be at risk due to  another attack surface emerging  in the process of implementation.} This is followed by a series of next questions, and in the course of providing answers, variants of the classic zero-dynamics attack are presented, from which the vulnerability of the target system is explored in depth. At the end, countermeasures are proposed to render the attack ineffective. Because it is known that the zero-dynamics in continuous-time systems cannot be modified by feedback, the main idea of the countermeasure is to relocate any unstable zero to a stable region in the stage of digital implementation through modified digital samplers and holders. Adversaries can still attack actuators, but due to the re-located zeros, they are of little use in damaging the target system.
}


\setcounter{page}{1}

\section{Introduction}\label{sec:1}

Modern control systems embrace the technology advances in communication and computing in order to increase their capability and applicability: feedback control systems now handle large scale systems whose subsystems are not necessarily collocated, and play key roles in societal infrastructures such as electrical power distribution facilities, oil and gas pipelines, water distribution systems and sewage treatment plants. The wide usage of Supervisory Control And Data Acquisition (SCADA) systems  to monitor and control operations of the large scale plant is an example. However, what also increased along with the capability and applicability is the {\em vulnerability} of the systems. Specifically, technological advances over the past decade have seen these traditionally closed systems become open and internet-connected, which puts the system at risk from attacks with malicious intent. Incidents of Iranian nuclear facility \cite{stuxnet}, breach in Maroochy water sewage \cite{maroochywater}, massive outage in Ukrainian power grid \cite{ukrainianpowergrid} show that the risk is not a possibility but a reality. Clearly, the effect of malfunction induced by malicious attack on control systems could be catastrophic as witnessed in \cite{droneiran, ukrainianpowergrid, stuxnet, Triton, maroochywater}. Naturally, security of control systems has become an active topic of research due to its importance.  

\begin{figure}[b]
    \centering
    \includegraphics[width=0.9\textwidth]{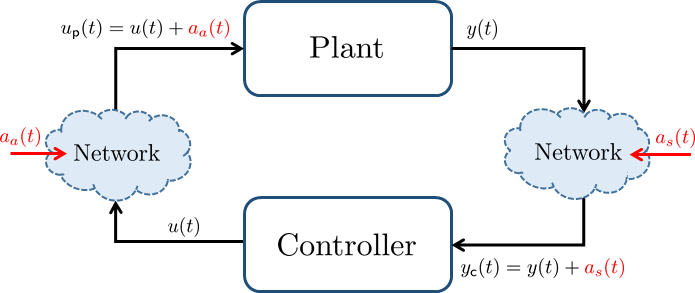}
    \caption{Illustration of modern control systems with remote communication under a sensor attack and an actuator attack}
    \label{fig:modern_cs}
\end{figure}

Figure \ref{fig:modern_cs} shows the modern control systems under consideration, where remotely located sensors and actuators are connected to computing units that execute feedback control algorithms, monitoring of the operation, and a level of anomaly detection. Here, values of sensor measurement, $y(t)$, are sent through the communication network and so are those for the actuators, $u(t)$. Attacks on sensors, denoted by $a_s(t)$, potentially corrupt sensor measurement $y(t)$ and so do the attacks on actuators, denoted by $a_a(t)$, the actuation signals for the plant.  

Community of automatic control began to investigate not only countermeasures to malicious attack but also mechanisms of various potential attacks on sensors, actuators, and controllers themselves \cite{park_stealthiness_2018,park_stealthy_2019,park_when_2016, lee_redundant_2019,lee_fully_2020}. What appears to be particularly dangerous is a class of attacks that are referred to be as {\em stealthy}. An attack being stealthy means that its effect does not appear in the output of the system or the output of the anomaly detector which is often deployed for the purpose of attack detection. A most well known one in this class is zero-dynamics attack \cite{teixeira_revealing_2012}: it is an actuator attack that pushes the system states along the zero dynamics. Due to the intrinsic property of zero dynamics, no sign shows in the output, which makes this attack stealthy. Another notion that we would like to bring up is an attack being {\em disruptive}. This means that a portion of the system states diverges due to the attack. A zero-dynamics attack is disruptive  if the associated zero dynamics is unstable. 

The central topic of this chapter is on the attacks that exploit the zero dynamics and also their countermeasures. Specifically, it is shown that zero-dynamics attack is feasible even without precise knowledge of the target systems (which is not the case in \cite{teixeira_revealing_2012}). Also shown is that it may be too soon to conclude that the systems having no zero dynamics are unsusceptible to such attacks: Sampling may bring a new vulnerability, and unstable pole dynamics may pose a similar threat if sensor attacks (as opposed to actuator attacks) are considered. Finally, we discuss zero-dynamics attack for nonlinear systems. As potential countermeasures, we introduce methods of moving zeros by generalized hold or by generalized sampling. Two methods move unstable zeros, that may be targets for the adversaries, to stable region in the complex plane. Then, although the systems can still be under stealthy attack, the attacks lose effectiveness and the adversaries lose motivation. All the results described are illustrated with numerical examples to help readers grasp the concepts.

\section{Threats of Zero-dynamics Attack}

Among many potential threats to the control systems, zero-dynamics attack is known to be lethal because it is fundamentally difficult to detect the attack.
Detection of the attack often takes place at the controller side, and so, the information available for detection is the input data transmitted from the controller and the output data received from the plant, both of which may be compromised over the communication networks.
Based on this fact, the attacker deceives the controller in a way that the transmitted input ($u$ in Figure \ref{fig:modern_cs}) and the received output ($y_\received = y+a_s$ in Figure \ref{fig:modern_cs}) would appear to be consistent with the dynamics  of the plant.

From attacker's perspective, attack specification consists of initiation time $t_0 > 0$, hazard level $L_\hazard$, and detection level $L_\detect$.
With them, it is said that the attack is {\em disruptive} if there exists $t^* \ge t_0$ such that
\begin{equation}\label{eq:disruptive}
 \|x(t^*) - x_\free(t^*)\| \ge L_\hazard
\end{equation}
where $x$ is the actual plant's state and $x_\free$ is the attack-free state that would have been resulted if there were no attack.
At the same time, the attack should be {\em stealthy}, which means that 
\begin{equation}\label{eq:stealthy}
\|y_\received(t) - y_\free(t) \| \le L_\detect \quad \text{for all $t$ such that $t_0 \le t \le  t^*$}
\end{equation}
where $y_\received$ is the received output and $y_\free$ is the attack-free output that would have been resulted if there were no attack.
Violation of \eqref{eq:stealthy} indicates that the system is under attack, and the level of detection $L_\detect$ is considered to be set by the defender, which is not to be zero for avoiding false alarm due to measurement noise and other perturbation.
In practice, rather than $y_\free$ (which is not simple to compute), the estimate $\hat y$ of $y$ is often used for attack detection where $\hat y$ is computed by the controller using the history of the input (sent to the plant) and the output (received from the plant) with the system model.
The idea behind the replacement of $y_\free$ with $\hat y$ is the following: If the estimator (that generates the estimate $\hat y$) is continuous with respect to its input $y_\received$ (for example, a dynamic system such as Luenberger observer), then (a) $\hat y(t) = y_\free(t)$ if the input to the estimator is $y_\free$; (b) $\hat y(t) \approx y_\free(t)$ if $y_\received(t) \approx y_\free(t)$ by continuity; and (c) $\|y_\received(t)-\hat y(t)\| \le \|y_\received(t)-y_\free(t)\|+\|y_\free(t)-\hat y(t)\|$, so that an attack that yields the compromised output $y_\received(t)$ arbitrarily close to $y_\free(t)$ can make $\|y_\received(t)-\hat y(t)\|$ arbitrarily small and thus is stealthy. 

In this section, the goal of attacker is to design an attack signal that is both disruptive and stealthy. We restrict the focus on attacks that exploit zero dynamics of the target systems.

\subsection{Zero-dynamics Attack}\label{sec:a}

Introduction of the zero-dynamics attack dates back at least to \cite{teixeira_revealing_2012}, where a geometric notion is employed to illustrate the idea of the zero-dynamics attack.
In this subsection, the zero-dynamics attack is re-interpreted in view of a canonical form that is a special realization of an LTI system.

Consider a continuous-time SISO LTI system represented by the following transfer function of relative degree $r$:
$$G(s) = \frac{{\beta_{n-r} s^{n-r} + \cdots + \beta_1 s + \beta_0}}{s^n+\alpha_{n-1} s^{n-1} + \cdots + \alpha_0}.$$	
It is known (from, e.g., \cite[Example 13.4]{khalil2014nonlinear}) that a state-space minimal realization of $G(s)$, called the (Byrnes-Isidori) normal form, is given by
\begin{align*}
y = x_1 \qquad \dot x_1 &= x_2 \\
\dot x_2 &= x_3 \\
&\vdots \\
\dot x_{r-1} &= x_r   \\
\dot x_r &= \phi_:^\top x_: + \phi_z^\top x_z + b u \\
\dot x_z &= S x_z + p x_1
\end{align*}
where $y$ is the output, $u$ is the input, and the state $x$ consists of $x_: = [x_1,x_2,\cdots,x_r]^\top \in \R^r$ and $x_z \in \R^{n-r}$.
The system parameters are $\phi_: \in \R^r$, $\phi_z \in \R^{n-r}$, $S \in \R^{(n-r) \times (n-r)}$, $p \in \R^{(n-r) \times 1}$, and $b \in \R$.
Here, it can be shown that the roots of zero polynomial ${\beta_{n-r} s^{n-r} + \cdots + \beta_1 s + \beta_0}$ coincide with the eigenvalues of the matrix $S \in \R^{(n-r) \times (n-r)}$ (see \cite[Example 13.4]{khalil2014nonlinear}).

\begin{svgraybox}
{\bf About zero dynamics:} 
Zero dynamics of a system is the residual sub-dynamics that remains when the output maintains identically zero by a pair of some initial condition and some input.
A benefit of the normal form is that the zero dynamics is easily identified, which is simply $\dot x_z = Sx_z$.
The pair of initial condition and input that make the output identically zero, is also identified as $(x_:(0),x_z(0)) = (0,\Delta)$ and $u(t) = -(1/b)\phi_z^\top x_z(t)$ where $\Delta$ is arbitrary.
Rigorously speaking, the dynamics $\dot x_z = S x_z + p x_1$ is called `internal dynamics,' but is sometimes called `zero dynamics' by abuse.
\end{svgraybox}

Suppose that an LTI dynamic controller $C(s) = H(sI-F)^{-1}G$ internally stabilizes the closed-loop system, and that an adversary can inject malicious signal $a_a(t)$ (called `actuator attack') into the input channel.
Then, the closed-loop system with the controller can be written as\footnote{We use the notation $\ourA$ and $\ourB$ (of suitable size) as
$$\ourA = \begin{bmatrix} 0 & 1 & 0 & \cdots & 0 \\ 0 & 0 & 1 & \cdots & 0 \\ \vdots & \vdots & \vdots & \ddots & \vdots \\
0 & 0 & 0 & \cdots & 1 \\ 0 & 0 & 0 & \cdots & 0 \end{bmatrix} \quad \text{and} \quad \ourB = \begin{bmatrix} 0 \\ 0 \\ \vdots \\ 0 \\ 1 \end{bmatrix} .$$}
\begin{align}\label{eq:closedloop}
    \begin{split}
    y = x_1 \qquad 
    \dot x_: &= \ourA x_: + \ourB (\phi_:^\top x_: + \phi_z^\top x_z + b(Hc+a_a)) \\
    \dot x_z &= Sx_z + p x_1 \\
    \dot c &= Fc + G x_1 
    \end{split}
\end{align}
where $c$ is the state of the controller.
Under the stability assumption, the following inequality holds for the system in (\ref{eq:closedloop}):
\begin{equation}\label{eq:converge1}
\left\| \begin{bmatrix} x_:(t) \\ x_z(t) \\ c(t) \end{bmatrix} \right\| \le k e^{-\lambda t} \left\| \begin{bmatrix} x_:(0) \\ x_z(0) \\ c(0) \end{bmatrix} \right\|
\end{equation}
with some positives $k$ and $\lambda$.

The following proposition is another interpretation of \cite{teixeira_revealing_2012} in terms of the normal form.

\begin{proposition}\label{prop:1}
Suppose that an actuator attack signal $a_a$ is generated by
\begin{equation}\label{eq:attack1}
    \dot z_\attack = S z_\attack, \qquad a_a = -\frac1b \phi_z^\top z_\attack
\end{equation}
with an initial condition $z_\attack(t_0) = \Delta$, and is injected from the time $t_0 > 0$ into the closed-loop.
If $\|\Delta\|$ is sufficiently small, then the attack becomes stealthy.
Moreover, if $S$ has at least one eigenvalue whose real part is positive (i.e., the plant has an unstable zero, or, is non-minimum phase) and the initial condition $\Delta$ excites the unstable mode, then the attack is disruptive.
\end{proposition}

\begin{proof}
While the attack generator is valid only for $t \ge t_0$, it can be equivalently rewritten as $\dot z_\attack = S z_\attack + \Delta \delta(t-t_0)$ with $z_\attack(0)=0$, which is valid for $t \ge 0$, where $\delta$ is Dirac's delta function.
Then, with $\tilde x_z := x_z - z_\attack$, the closed-loop system can be rewritten as
\begin{align}\label{eq:underattack}
    \begin{split}
    y = x_1 \qquad 
    \dot x_: &= \ourA x_: + \ourB (\phi_:^\top x_: + \phi_z^\top \tilde x_z + bHc) \\
    \dot {\tilde x}_z &= S \tilde x_z + p x_1 - \Delta \delta(t-t_0) \\
    \dot c &= Fc + G x_1 .
    \end{split}
\end{align}
By comparing \eqref{eq:closedloop} (without $a_a$) and \eqref{eq:underattack}, it is seen that they are the same except the replacement of $x_z$ with $\tilde x_z$ and the perturbation $\Delta$.
The perturbation causes a transient response after $t_0$.
Because the attack-free response $x_\free$ in \eqref{eq:stealthy} is the same as the response of \eqref{eq:underattack} with $\Delta = 0$, and because the system \eqref{eq:underattack} is stable, it is seen that $y(t)$ (which is the same as $y_\received(t)$ since there is no sensor attack) can be made arbitrarily close to $y_\free(t)$ (which is the first component of $x_\free(t)$) by choosing $\|\Delta\|$ small enough. 
Hence, the attack is stealthy (for any $t^*$ in \eqref{eq:stealthy}).
Now, by choosing a small $\Delta$ that excites the unstable mode of \eqref{eq:attack1}, the state $x_z(t)$ diverges because $z_\attack(t)$ diverges while $\tilde x_z(t) = x_z(t)-z_\attack(t)$ exponentially converges to zero.
This shows the attack is disruptive.
\qed
\end{proof}

\begin{remark}
If $S$ is expressed in a coordinate where $S = {\rm blockdiag}(S_\uu,S_\sss)$ where all the eigenvalues of $S_\sss$ have positive real parts, then the dimension of attack generator \eqref{eq:attack1} can be reduced as $\dot z_\attack = S_\sss z_\attack$ and $a_a = -(1/b) \phi_{z,\sss}^\top z_\attack$ where $\phi_{z,\sss}$ is the corresponding part of $\phi_z$.
\end{remark}

\subsubsection{Pole-dynamics Attack}

As a variation of the zero-dynamics attack, which is an actuator attack, we consider its dual `pole-dynamics attack' that is a sensor attack studied in \cite{jeon_resilient_2016}.
If we rewrite the controller using the normal form, the closed-loop system is written by
\begin{align}\label{eq:controller1}
\begin{split}
    u = c_1 \qquad \dot c_: &= \ourA c_: + \ourB ( \phi_:^\top c_: + \phi_z^\top c_z + b (y + a_s) ) \\
    \dot c_z &= S c_z + p c_1 \\
    \dot x &= A x + B c_1, \qquad y = C x
\end{split}
\end{align}
where $x$ is the state of the plant $C(sI-A)^{-1}B$, and $\phi_:$, $\phi_z$, $b$, $S$, and $p$ are the parameters of the controller.
By comparison, it is seen that the same argument holds for the design of the sensor attack $a_s$.

\begin{proposition}
Suppose that a sensor attack signal $a_s$ is generated by
\begin{equation}
    \dot z_\attack = A z_\attack, \qquad a_s = - C z_\attack
\end{equation}
with an initial condition $z_\attack(t_0) = \Delta$, and is injected from the time $t_0 > 0$ into the closed-loop.
If $\|\Delta\|$ is sufficiently small, then the attack becomes stealthy.
Moreover, if $A$ has at least one eigenvalue whose real part is positive (i.e., the plant has an unstable pole, or the plant is unstable) and the initial condition $\Delta$ excites the unstable mode, then the attack is disruptive.
\end{proposition}

\begin{proof}
With $\tilde x := x - z_\attack$, we have
\begin{align}\label{eq:controller2}
\begin{split}
    u = c_1 \qquad \dot c_: &= \ourA c_: + \ourB ( \phi_:^\top c_: + \phi_z^\top c_z + b C \tilde x ) \\
    \dot c_z &= S c_z + p c_1 \\
    \dot {\tilde x} &= A \tilde x + B c_1 - \Delta \delta(t-t_0)
\end{split}
\end{align}
so that $(x(t)-z_\attack(t))$ exponentially converges to zero after $t_0$, so that $x(t)$ diverges if $\Delta$ excites unstable mode of $A$. Hence, the attack is disruptive.
At the same time, with sufficiently small $\Delta$, the states of \eqref{eq:controller2} can be made arbitrarily close to the states of \eqref{eq:controller1} without $a_s$.
Therefore, the received output $y_\received = y - a_s = Cx - Cz_\attack = C\tilde x$ can be made arbitrarily close to the output $y$ of \eqref{eq:controller1} without $a_s$ which corresponds to $y_\free$. Hence, the attack is stealthy by \eqref{eq:stealthy}.
\qed
\end{proof}

The zero-dynamics attack is a stealthy attack regardless whether the plant has unstable zeros or not.
However, the attack becomes disruptive when there is an unstable zero.
At this moment, a natural question arises: ``Does it mean that the control system is secure from disruptive stealthy attacks if the plant has no unstable zeros?''

\subsection{Sampling-zero-dynamics Attack}

\begin{figure}
  \centering
  \includegraphics[width=0.9\textwidth]{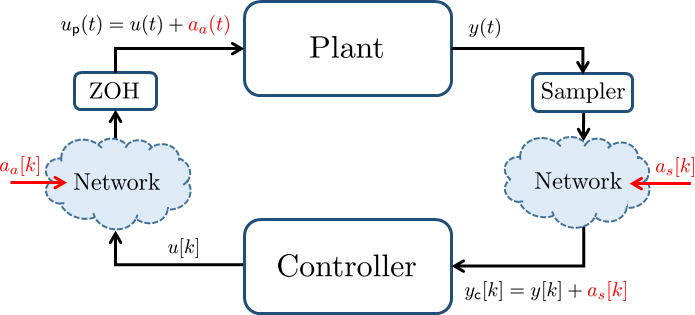}
  \caption{Control system implemented with zero-order hold and sampler}
  \label{fig:v1}
\end{figure}

Unfortunately, a disruptive and stealthy  attack that exploits zero dynamics can be designed  even if there is no unstable zeros {\em in continuous-time}.
Modern control systems are mostly implemented by a digital controller, and thus, the continuous-time output signal of the plant is sampled to a discrete-time signal, and the discrete-time control input is converted to a continuous-time signal by zero-order hold (see Fig.~\ref{fig:v1}).
The role of sampler and zero-order hold is defined as
\begin{itemize}
\item sampler: $y[k] := y(kT_s)$
\item zero-order hold: $u(t) = u[k]$, $kT_s \leq t < (k+1)T_s$
\end{itemize}
where $T_s$ is the sampling period and $k$ is the discrete-time index.

Suppose that the continuous-time plant is given by $G(s)$ whose relative degree is~$r$.
Then, representation of the plant from  the discrete-time input $u[k]$ to output $y[k]$ becomes the discrete-time transfer function $\bar G(z)$.
It is emphasized that the relative degree $\bar r$ of $\bar G(z)$ is equal to 1 for almost all sampling periods $T_s$, regardless of the relative degree $r$ of $G(s)$.
This means that sampling introduces $r-1$ additional zeros, and these new zeros are called {\em sampling zeros}.
The bad news is that, if $r \ge 3$ and the sampling is fast so that $T_s$ is small, then at least one of the sampling zeros is always unstable, and so, the attacker can deploy the {\em sampling-zero-dynamics attack} even if the continuous-time plant has no unstable zeros.

\begin{svgraybox}
\noindent{\bf About sampling zeros:}
Let us first ask why the discrete-time plant $\bar G(z)$ has $r-1$ additional zeros over its continuous-time counterpart $G(s)$ for almost all $T_s$.
To see this, let us examine a typical step response of, e.g., $G(s) = (-s+3)/(s^3+3s^2+3s+1)$ as follows:
\begin{center}
\includegraphics[width=0.55\textwidth]{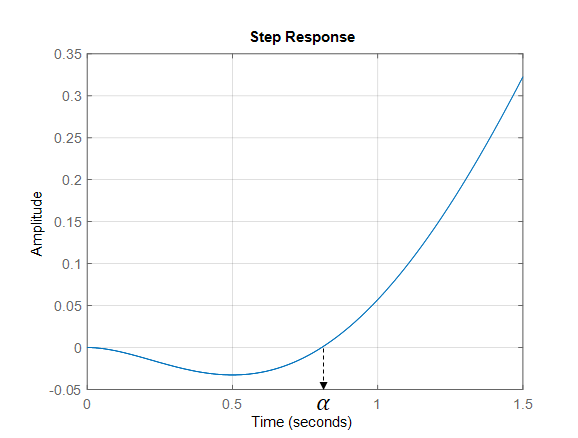}
\end{center}

Since the discrete-time step input $u[k]$ is converted as a unit step continuous-time input $u(t)$, the discrete-time step response $y[k]$ can be found from the plot, that is, $y[k] = y(k T_s)$.
Then, from the plot, it is seen that $y[1] = y(T_s) \not = 0$ for almost all $T_s$, i.e., except $T_s = \alpha$ in the plot.
This means that the relative degree of $\bar G(z)$ is 1 except the case $T_s = \alpha$, even if the relative degree of $G(s)$ is $r=2$.
Thus, there should be one sampling zero in $\bar G(z)$.

In fact, the following formal statement is taken from \cite{YuzGoodwin2014Book}:
\begin{lemma}
For almost all $T_s>0$,
$$\bar G(z) = g \frac{\gamma_{r-1}(z) \prod_{i=1}^{n-r} (z - \bar z_i)}{\prod_{i=1}^{n}(z - \bar p_i)}$$
where 
\begin{itemize}
    \item $\bar p_i = e^{\lambda_i T_s}$ is the pole where $\lambda_i$ is the pole of $G(s)$
    \item $\bar z_i$ is the zero (When $T_s$ is sufficiently small, $\bar z_i$ gets close to $e^{\mu_i T_s}$ where $\mu_i$ is the zero of $G(s)$. Since these zeros appear from the continuous-time zeros, they are called `intrinsic zeros.')
    \item $\gamma_{r-1}(z)$ is a polynomial of order $r-1$, dependent on the sampling time $T_s$ (As $T_s \to 0$, the coefficients of $\gamma_{r-1}$ converge to those of the Euler-Frobenius polynomial $P_{r-1}(z)$ of order $r-1$. The roots of $\gamma_{r-1}(z)$ are called `sampling zeros.')
    \item $g$ is a constant defined by $(T_s)^r (\lim_{s \to \infty}s^r G(s)) / r!$.
\end{itemize}
\end{lemma}
The Euler-Frobenius polynomial of order $r-1$ is defined as $P_{r-1}(z) = \beta_{r-1} z^{r-1} + \beta_{r-2} z^{r-2} + \cdots + \beta_1 z + \beta_0$ where $\beta_i = \sum_{j=0}^i (-1)^{i-j}(j+1)^{r}\binom{r+1}{i-j}$, which is known to have at least one unstable root that is located outside of the unit circle in the complex plane whenever $r \ge 3$ \cite{YuzGoodwin2014Book}.
Therefore, if the continuous-time plant $G(s)$ has relative degree $r \ge 3$ and the sampling period $T_s$ is sufficiently small, then appearance of unstable sampling-zero is unavoidable.
All the roots of Euler-Frobenius polynomial are negative real.
An implication is that the output-zeroing input corresponding to the sampling zero should have alternating signs.
\end{svgraybox}

Since the relative degree of $\bar G(z)$ is 1 for almost all cases, let us suppose that $\bar G(z)$ is realized in the (discrete-time) normal form:
\begin{align}\label{eq:dtnormal}
\begin{split}
    y[k] = x_1[k] \qquad x_1[k+1] &= \phi_: x_1[k] + \phi_z^\top x_z[k] + b (u[k] + a_a[k]) \\
    x_z[k+1] &= S x_z[k] + p x_1[k] 
\end{split}
\end{align}
where $x_z \in \R^{n-1}$.

\begin{proposition}
Suppose that an actuator attack signal $a_a[k]$ is generated by
\begin{equation}\label{eq:dtattack}
    z_\attack[k+1] = S z_\attack[k], \qquad a_a[k] = -\frac{1}{b} \phi_z^\top z_\attack[k]
\end{equation}
with $z_\attack[k_0] = \Delta$ where the attack initiates at $t_0 = k_0 T_s$.
If $\|\Delta\|$ is sufficiently small, then the attack becomes stealthy.
Moreover, if $S$ has at least one eigenvalue outside of the unit circle (which is either an instrinsic zero or a sampling zero) and the initial condition $\Delta$ excites the unstable mode, then the atack is disruptive.
\end{proposition}

\begin{proof}
Define $\tilde x_z := x - z_\attack$ and consider the controller is given by $\bar C(z) = H(zI-F)^{-1}G$.
Then, the closed-loop system becomes
\begin{align}
\begin{split}
    y[k] = x_1[k] \qquad x_1[k+1] &= \phi_: x_1[k] + \phi_z^\top \tilde x_z[k] + bHc[k] \\
    \tilde x_z[k+1] &= S \tilde x_z[k] + p x_1[k]  - \Delta \delta[k-k_0] \\
    c[k+1] &= Fc[k] + Gx_1[k]
\end{split}
\end{align}
where $\delta$ is Kronecker's delta.
The rest of argument is similar to the proof of Proposition \ref{prop:1}.
\qed
\end{proof}

\begin{example}\label{ex:2Mass}

\begin{figure}[t]
\sidecaption
\includegraphics[width=0.6\textwidth]{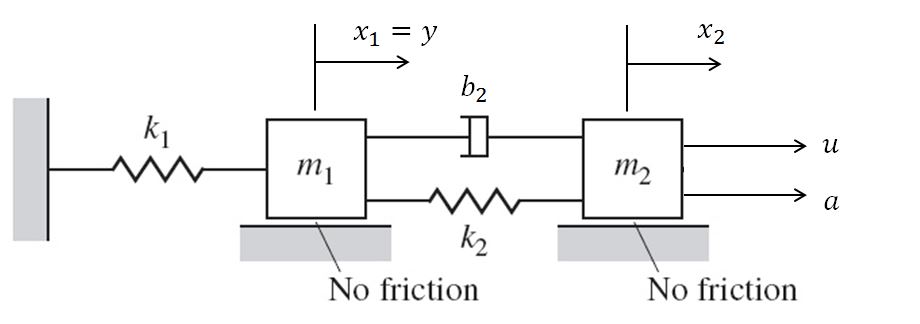}
\caption{Two mass system with all parameters are set to 1.
(Reprinted from \cite{kim_neutralizing_2020}, Copyright 2020, with
permission from Elsevier)}
\label{fig:v2}
\end{figure}

\begin{figure}[b]
\includegraphics[width=0.8\textwidth]{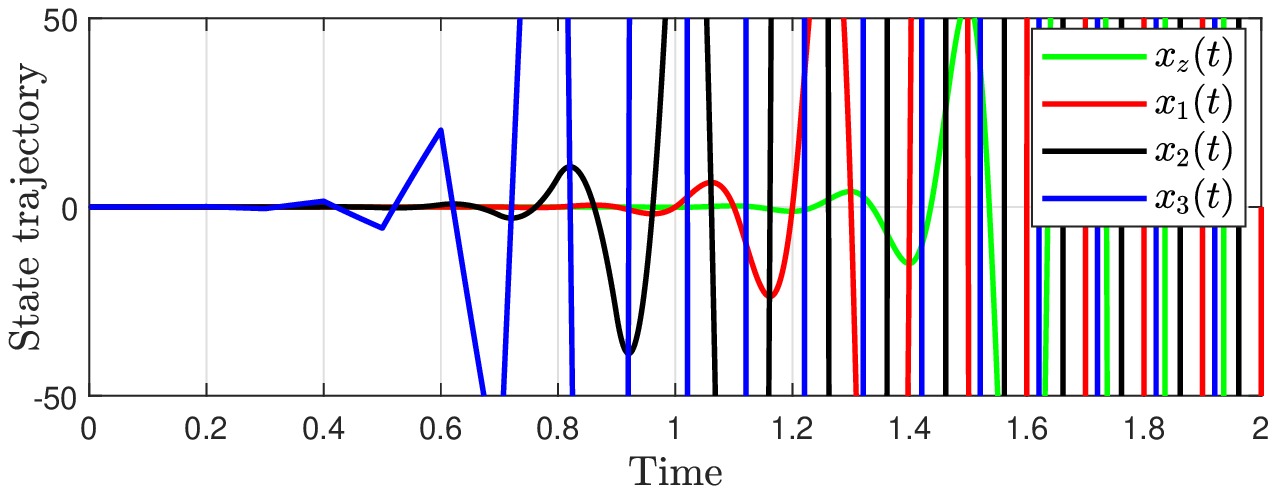} \\
\includegraphics[width=0.8\textwidth]{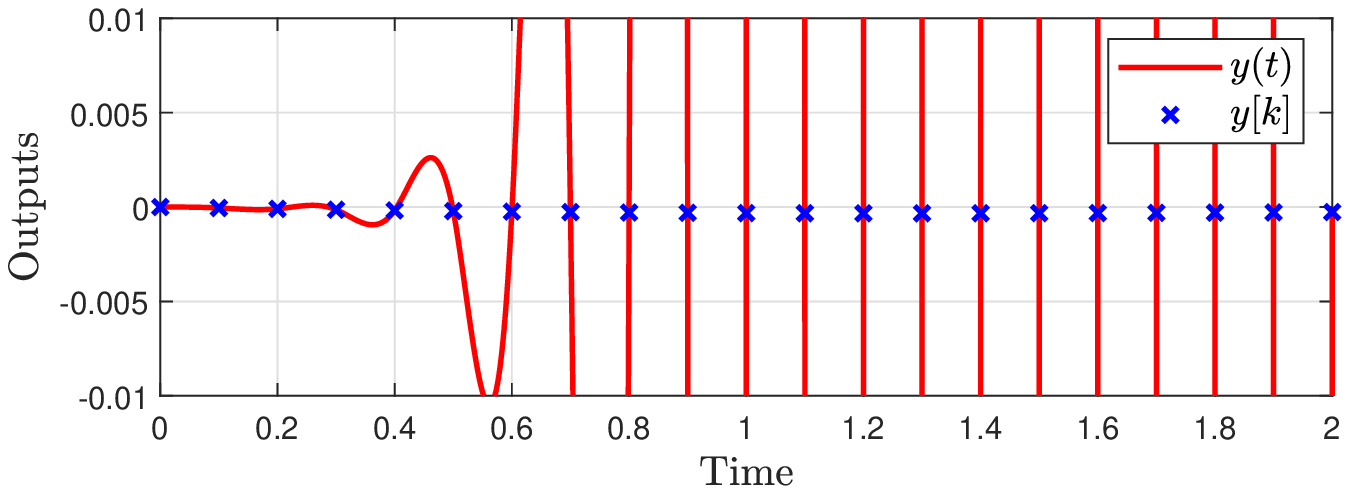}
\caption{The first plot is the state trajectory and the second is the outputs. Red color is the output $y(t)$, and the sampled output $y[k]$ is marked as blue cross.
(Reprinted from \cite{kim_neutralizing_2020}, Copyright 2020, with
permission from Elsevier)}
\label{fig:v3}
\end{figure}

The transfer function of Fig.~\ref{fig:v2}  from $u(t)$ to $y(t)=x_1(t)$ is $G(s) = (s+1)/(s^4+2s^3+3s^2+s+1)$, which is a minimum phase system having relative degree~3.
With $T_s = 0.1$, its discrete-time equivalent model can be written as \eqref{eq:dtnormal} with
\begin{gather*}
    S = \begin{bmatrix} 0 & 1 & 0 \\ 0 & 0 & 1 \\ 0.86 & 2.58 & -2.99 \end{bmatrix},
    \quad p = \begin{bmatrix} 0 \\ 0 \\ 1 \end{bmatrix}, 
    \quad \phi_z = \begin{bmatrix} 5.02 \\ 20.04 \\ -28.3 \end{bmatrix} \\
    \phi_: = 6.78, \quad b = 1.62 \times 10^{-4}.
\end{gather*}
Note that the eigenvalues of $S$ are $-3.64$, $-0.26$, and $0.90$, of which the first two are sampling zeros. 
Clearly, there is an unstable zero of $-3.64$.
With $\Delta = 10^{-5}\times[1, 0, 0]^T$, the attack generated by \eqref{eq:dtattack} results in Fig.~\ref{fig:v3}.
\qed
\end{example}

Up to now, we have seen that even if the continuous-time plant does not have unstable zero, sampling for discrete-time implementation of control system may introduce unstable sampling zeros in discrete-time domain.
At this moment, a natural question arises: ``If the discrete-time plant has no unstable (sampling) zero, does it mean that the control system is secure from disruptive stealthy attacks?''

\subsection{Enforced-zero-dynamics Attack}

\begin{figure}[b]
\sidecaption
\includegraphics[width=0.6\textwidth]{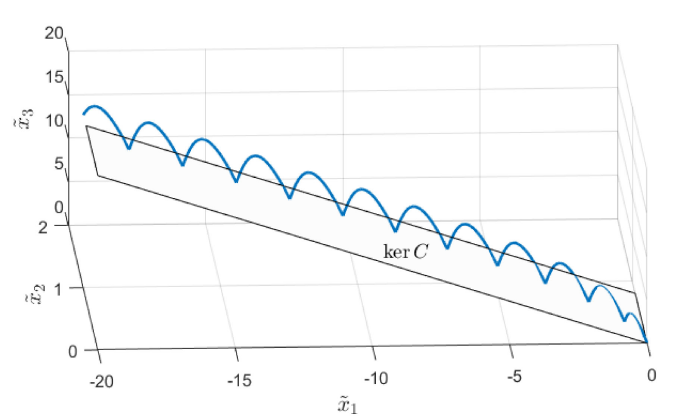}
\caption{Solution diverges from the origin while it belongs to the kernel of the output matrix $C$ at each sampling instance.
(©2020 IET. Reprinted, with permission, from \cite{kim_masking_2019})}
\label{fig:v7}
\end{figure}

Unfortunately, even if the plant has no unstable zero in the discrete-time domain, there is some possibility to design a disruptive stealthy attack when there is more freedom in actuation than sensing.
Since sensing occurs only at discrete time instances, if an actuator attack drives the plant's state to pass through the kernel of the output matrix at each sensing time while the state is enforced to behave disruptively, then the attack can remain stealthy while the state diverges (see Fig.~\ref{fig:v7} for example).
This type of attack exists when the sampled-data system has an input redundancy, i.e., the number of inputs being larger than that of outputs and/or the sampling rate of the actuators being higher than that of the sensors. 
For the sake of brevity, only the latter case is explained in this chapter in detail. 

Consider a linear time-invariant plant
\begin{align}\label{eq:plant23}
\begin{split}
    \dot x(t) &= A x(t) + B (u(t)+a_a(t)) \quad \in \R^n, \qquad\qquad u(t), \; a_a(t) \quad \in \R^p, \\
    y(t) &= C x(t) \quad \in \R^q
\end{split}
\end{align}
where $a_a$ is the actuator attack, and the matrices $B$ and $C$ are assumed to have full column rank and full row rank, respectively.
Let $T_s$ be the sampling period for the output measurement such that 
$$y[k] = y(kT_s)$$
and $T_a$ be the period for zero-order-hold such that
$$u(t) + a_a(t) = u[i] + a_a[i], \quad iT_a \le t < (i+1)T_a$$
in which, it is assumed that a discrete-time attack $a_a[i]$ is injected in the communication network so that it is converted together with $u[i]$.
Then, the enforced-zero-dynamics attack $a_a[i]$ can be generated under the assumption that
\begin{equation}\label{eq:freedom}
qT_a < p T_s .
\end{equation}
It should be noted that \eqref{eq:freedom} holds if the number of inputs, $p$, is large and/or the sampling period for actuation, $T_a$, is small.
Based on \eqref{eq:freedom}, a disruptive stealthy attack is proposed in \cite{kim_masking_2019}, and the attack is termed as `masking attack.'

In order to introduce the idea of the masking attack (we call it `enforced-zero-dynamics attack' in this chapter), let us simplify the situation as
\begin{equation}\label{eq:freedom2}
p = q = 1, \qquad T_s = 2 T_a
\end{equation}
and refer to \cite{kim_masking_2019} for the general cases.
Now, let 
$$A_\ddd := e^{A T_a} \quad \text{and} \quad B_\ddd := \int_{0}^{T_a} e^{A\tau} d\tau \; B .$$
Also, let $\tilde x := x - x_\free$ and $\tilde y := y - Cx_\free$ where $x_\free$ is the solution of \eqref{eq:plant23} when there is no attack, and let the attack initiate at $t_0=0$ (so that $\tilde x(0) = 0$).
Then, it is seen that 
\begin{subequations}\label{eq:aa}
\begin{align}
\begin{bmatrix} \tilde x(T_a) \\ \tilde x(2T_a) \end{bmatrix} &= \begin{bmatrix} B_\ddd & 0 \\ A_\ddd B_\ddd & B_\ddd \end{bmatrix} \begin{bmatrix} a_a[0] \\ a_a[1] \end{bmatrix} \label{eq:aa1} \\
\tilde y[1] = \tilde y(T_s) = \tilde y(2T_a) &= \begin{bmatrix} C A_\ddd B_\ddd & C B_\ddd \end{bmatrix} \begin{bmatrix} a_a[0] \\ a_a[1] \end{bmatrix} . \label{eq:aa2}
\end{align}
\end{subequations}

Now, we assume that
\begin{flalign*}
\text{(a)} &\; \ker \begin{bmatrix} C A_\ddd B_\ddd, & C B_\ddd \end{bmatrix} \not \subset \ker \begin{bmatrix} B_\ddd & 0 \\ A_\ddd B_\ddd & B_\ddd \end{bmatrix}, & \\
\text{(b)} &\; {\rm im} \; C A_\ddd^2 \subset {\rm im} \; \begin{bmatrix} C A_\ddd B_\ddd, & C B_\ddd \end{bmatrix}, &
\end{flalign*}
which automatically follow for almost all sampling periods from the assumption \eqref{eq:freedom2} and that the matrix $B$ has full column rank.
Indeed, item (a) holds because $B_\ddd$ also has full column rank so that the right-hand is the trivial set while the left-hand is not trivial because the matrix is fat.
In addition, item (b) follows from the facts that the relative degree of $C(zI-A_\ddd)^{-1}B_\ddd$ is 1 for almost all sampling periods and that $CB_\ddd$ corresponds to the relative degree of the transfer function.
It is emphasized again that
\begin{flalign*}
\text{(c)} &\; \ker \begin{bmatrix} C A_\ddd B_\ddd, & C B_\ddd \end{bmatrix} \; \text{is not trivial.} &
\end{flalign*}

Now, it is ready to design an attack sequence $a_a[i]$.
The design proceeds sequentially as a pair; i.e., choosing the sequences $\alpha_m, \beta_m \in \R^2$ such that
\begin{equation}\label{eq:a_a}
\begin{bmatrix} a_a[2m] \\ a_a[2m+1] \end{bmatrix} = \beta_m + \alpha_m, \qquad m=0,1,\dots.
\end{equation}
For the case $m=0$, let $\beta_0=0$, and pick $\alpha_0$ such that 
$$\begin{bmatrix} C A_\ddd B_\ddd & C B_\ddd \end{bmatrix} \alpha_0 = 0 \quad \text{and} \quad 
\begin{bmatrix} B_\ddd & 0 \\ A_\ddd B_\ddd & B_\ddd \end{bmatrix} \alpha_0 \not = 0$$
which is possible by the items (a) and (c).
Then, by \eqref{eq:a_a} with $m=0$, it follows from \eqref{eq:aa} that $\tilde y[1]=0$ and at least one of $\tilde x(T_a)$ and $\tilde x(2T_a)$ is non-zero.
Moreover, by choosing $\alpha_0$ to have sufficiently large magnitude, the non-zero one can be made arbitrarily large in magnitude.
Therefore, both the stealthy property and the disruptive property of the attack are satisfied in the attack period of $2T_a$.

However, it may require too much actuation effort to achieve the level of hazard $L_\hazard$ in \eqref{eq:disruptive} within the period of $2T_a$.
In order to distribute the effort over time, let us design the attack \eqref{eq:a_a} for next steps $m \ge 1$.
Since
\begin{subequations}
\begin{align}
\begin{bmatrix} \tilde x((2m+1)T_a) \\ \tilde x(2(m+1)T_a) \end{bmatrix} &= 
\begin{bmatrix} A_\ddd \tilde x(2m T_a) \\ A_\ddd^2 \tilde x(2m T_a) \end{bmatrix} +
\begin{bmatrix} B_\ddd & 0 \\ A_\ddd B_\ddd & B_\ddd
\end{bmatrix} \begin{bmatrix} a_a[2m] \\ a_a[2m+1] \end{bmatrix} \notag \\
&= \begin{bmatrix} A_\ddd \tilde x(2m T_a) \\ A_\ddd^2 \tilde x(2m T_a) \end{bmatrix} +
\begin{bmatrix} B_\ddd & 0 \\ A_\ddd B_\ddd & B_\ddd
\end{bmatrix} \beta_m + \begin{bmatrix} B_\ddd & 0 \\ A_\ddd B_\ddd & B_\ddd
\end{bmatrix} \alpha_m \label{eq:cc1} \\
\tilde y[m+1] &= \tilde y((m+1)T_s) = C A_\ddd^2 \tilde x(2mT_a) + \begin{bmatrix} C A_\ddd B_\ddd & C B_\ddd \end{bmatrix} \begin{bmatrix} a_a[2m] \\ a_a[2m+1] \end{bmatrix} \notag \\
&= C A_\ddd^2 \tilde x(2mT_a) + \begin{bmatrix} C A_\ddd B_\ddd & C B_\ddd \end{bmatrix} \beta_m + \begin{bmatrix} C A_\ddd B_\ddd & C B_\ddd \end{bmatrix} \alpha_m , \label{eq:cc2}
\end{align}
\end{subequations}
the vector $\beta_m$ is chosen first such that
$$C A_\ddd^2 \tilde x(2mT_a) + \begin{bmatrix} C A_\ddd B_\ddd & C B_\ddd \end{bmatrix} \beta_m = 0$$
which is possible thanks to the item (b).
The role of $\beta_m$ is to cancel the effect of previous actuation in \eqref{eq:cc2} in order for the stealthiness of the step $m$.
Then, with $\beta_m$ at hand, the vector $\alpha_m$ is chosen in $\ker \begin{bmatrix} C A_\ddd B_\ddd, & C B_\ddd \end{bmatrix}$, so that $\tilde y[m+1]=0$ from \eqref{eq:cc2}, but not in the kernel of 
$$\begin{bmatrix} B_\ddd & 0 \\ A_\ddd B_\ddd & B_\ddd \end{bmatrix}$$
such that the states in \eqref{eq:cc1} can reach the level $L_\hazard$.
By repeating the procedure for $m=2,3,\cdots$, a disruptive stealthy attack can be generated.

It is noted that the attack generation can be performed off-line because it does not require any online information such as input and output signals.
Therefore, the attack sequence $a_a[i]$ can be computed and stored before the attack initiation.

\begin{figure}[t]
\sidecaption
\includegraphics[width=0.7\textwidth]{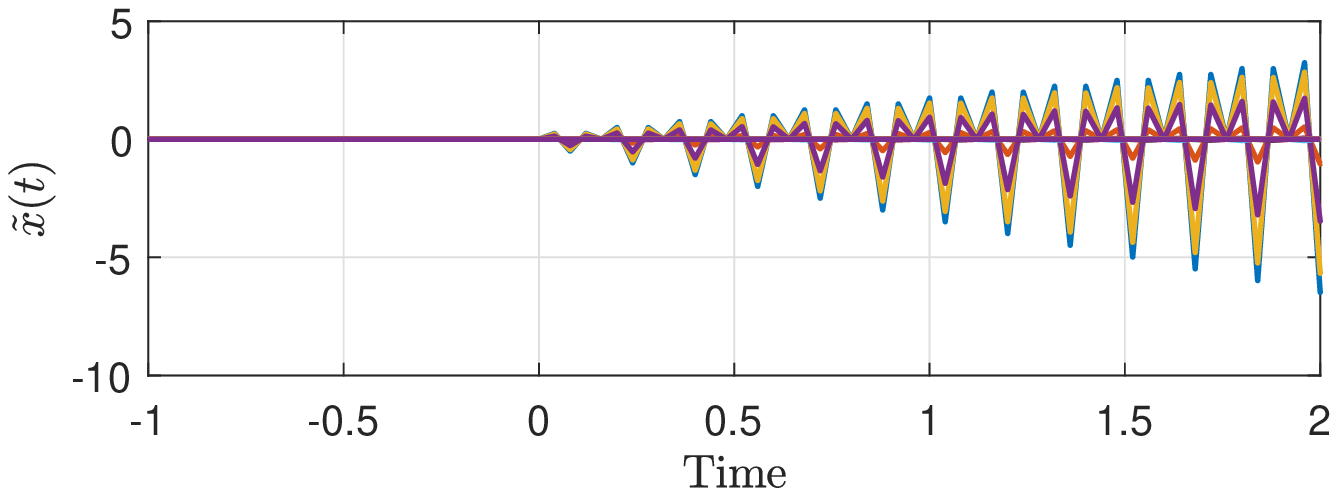} \\
\includegraphics[width=0.7\textwidth]{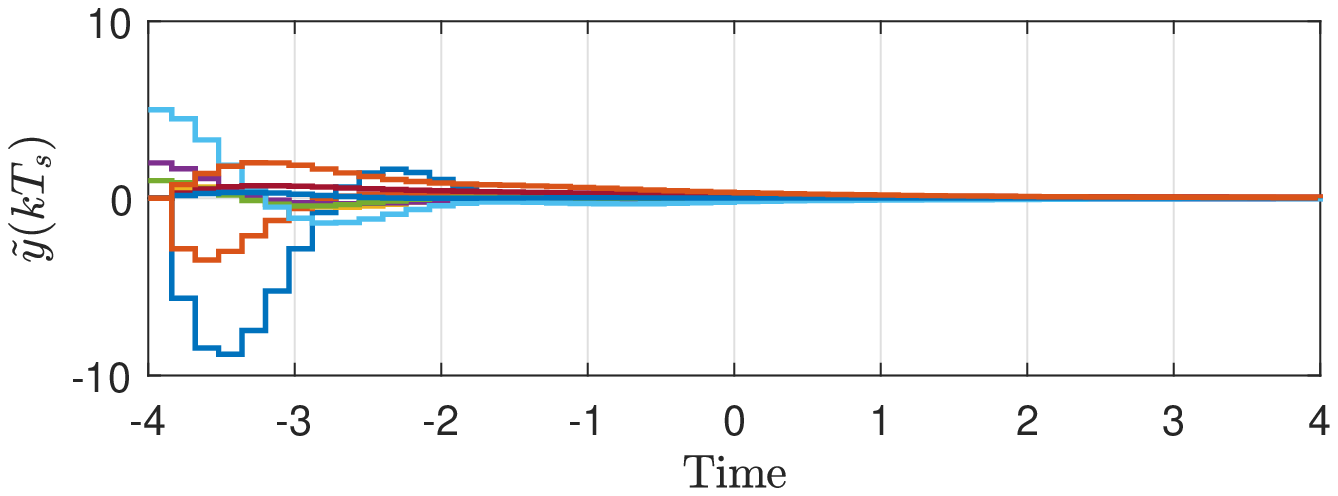}
\caption{The system reaches its steady-state before the time $t_0=0$, and enforced-zero-dynamics attack begins at $t_0=0$. While the sampled output $\tilde y(kT_s)$ remains identically zero, the continuous-time state deviation $\tilde x(t)$ is diverging.
(©2020 IET. Reprinted, with permission, from \cite{kim_masking_2019})}
\label{fig:v8}
\end{figure}

\begin{example}
Consider a model of X-38 vehicle in \cite{shieh2000design} which has 11 states, 3 inputs, and 9 outputs.
The system admits a multi-rate controller (see \cite{shieh2000design}), and we suppose that $T_s = 4 T_a$ so that the inequality \eqref{eq:freedom} holds.
Following the procedure of \cite{kim_masking_2019}, an actuator attack signal is generated off-line before the attack initiation.
Fig.~\ref{fig:v8} illustrates the outcome of the attack.
\qed
\end{example}

All of the attacks discussed so far require exact knowledge of the plant. 
In practice, exact model of the physical system is hard to obtain not only for attackers but also for control designers. 
At this moment, a natural question arises: ``It seems that these attack policies are too ideal. Is the real control system secure because exact model knowledge is hard to obtain?''

\subsection{Robust Zero-dynamics Attack}

Unfortunately, exact model of physical plant is not necessary for generating disruptive stealthy attacks, if the attack policy is designed by employing a robust control methodology.
In this subsection, let us investigate this possibility using a particular robust control, called {\em disturbance observer approach} \cite{shim2016yet}, and generate a disruptive stealthy actuator attack for continuous-time linear systems.

For this, let us write the plant in the normal form again:
\begin{equation}\label{eq:realplant}
\begin{aligned}
    y = x_1, \qquad 
    \dot x_: &= \ourA x_: + \ourB (\phi_:^\top x_: + \phi_z^\top x_z + b(u+a_a)), & x_: &\in \R^r, \\
    \dot x_z &= Sx_z + p x_1, & x_z &\in \R^{n-r} .
\end{aligned}
\end{equation}
While the proposed attack policy does not need exact values of the system parameters $\phi_:$, $\phi_z$, $b$, $S$, and $p$, some information regarding the plant is still needed.
They are:
\begin{itemize}
\item the input and the output of the plant: instead of using less knowledge on the plant model, the attacker relies more on the input and the output information (readers are reminded that the classical zero-dynamics attack does not use the output of the plant),
\item the relative degree $r$, and the sign of the high-frequency gain $b$, and nominal values $\phi_{:,\nn}$, $\phi_{z,\nn}$, $b_\nn$, $S_\nn$, $p_\nn$,
\item attack-free operating region $\XX$ and $\ZZ$ such that $x_:(t) \in \XX$ and $x_z(t) \in \ZZ$, $\forall t \ge 0$: these are the underlying sets for attack-free operation, and both $\XX$ and $\ZZ$ are assumed to be compact sets; they can be taken conservatively,
\item the amount of variation of uncertain parameters: for example, (possibly conservative) interval to which the uncertain parameter $b$ belongs; it is also assumed that the nominal values belong to the intervals.
\end{itemize}

The above information may be obtained by the attacker from a system identification procedure using the eavesdropped input and output signals, or from a leaked information about the design model that was used for constructing the feedback controller.
In any case, the amount of information required is not much different from the knowledge to design a robust controller for the uncertain plant.
Finally, for technical reason, we assume the following.

\smallskip
\noindent
{\bf Assumption.}
The (uncertain) matrix $S$ does not have eigenvalues on the imaginary axis in the complex plane.
\smallskip

To understand the idea of robust zero-dynamics attack, suppose, for the time being, that the attacker generates the actuator attack signal (instead of \eqref{eq:attack1}) by
\begin{subequations}\label{eq:rattack}
\begin{align}
    \dot z_\attack &= S_\nn z_\attack + p_\nn y \label{eq:rattack1} \\
    a_a^* &= \frac{1}{b} \left( -\phi_:^\top x_: - \phi_z^\top x_z - bu + \phi_{:,\nn}^\top x_: + \phi_{z,\nn}^\top z_\attack + b_\nn u \right). \label{eq:rattack2}
\end{align}
\end{subequations}
Of course, the signal $a_a^*$ is not available because many terms in the right-hand side are unknown.
This signal $a_a^*$ will be {\em estimated} by the disturbance observer later.
With $a_a = a_a^*$, the plant \eqref{eq:realplant} is rewritten (together with \eqref{eq:rattack1}) by
\begin{subequations}\label{eq:p}
\begin{align}
    y = x_1 \qquad 
    \dot x_: &= \ourA x_: + \ourB (\phi_{:,\nn}^\top x_: + \phi_{z,\nn}^\top z_\attack + b_\nn u) \label{eq:p1} \\
    \dot z_\attack &= S_\nn z_\attack + p_\nn x_1 \label{eq:p2} \\
    \dot x_z &= Sx_z + p x_1 . \label{eq:p3}
\end{align}
\end{subequations}
It is seen from the above equation that all the uncertain parameters are replaced with their nominal values.
In particular, the attacker's dynamics \eqref{eq:p2} replaces the role of the real zero dynamics \eqref{eq:p3}, and disguises as the plant's internal dynamics.
The controller interacts with the attacker's dynamics because the state $z_\attack$ is seen from the output $y$ while the state $x_z$ becomes unobservable, and thus, the real zero dynamics \eqref{eq:p3} is placed out of the feedback loop.

It is reasonable to assume that the controller robustly stabilizes the uncertain plant whose parameters belong to the uncertainty interval, which also contains the nominal parameters.
Then, since the controller stabilizes the plant \eqref{eq:p} with the replaced nominal parameters, the output $y$ and the state $(x_:,z_\attack)$ behave as if there is no attack.
(This means that the trajectory of $(x_:(t),z_\attack(t))$ corresponds to $x_\free(t)$ in \eqref{eq:stealthy}.)
In this way, the attack becomes stealthy.
On the other hand, if the uncertain plant has an unstable zero so that the matrix $S$ has an unstable eigenvalue (having positive real parts), then the state $x_z$ tends to diverge for {\em almost all} values of $x_z(t_0)$ where $t_0$ is the time of attack initiation.
In this way, the attack becomes disruptive.
It should be noted that the way of obtaining disruptive property is different from that of the zero-dynamics attack in Section \ref{sec:a}.

\begin{svgraybox}
{\bf Why not $x_z$ diverges for all $x_z(t_0)$?}
Even if a system is unstable, there is a measure-zero set of initial conditions from which the solution remains bounded.
For example, $x(0)=0$ for the unstable system $\dot x = x$ yields a bounded solution $x(t)=0$.
This is true even when the system has a bounded external input.
An example is $\dot x = x + \sin t$.
With $x(0) = -1/2$, the solution $x(t) = -(1/2)(\sin t + \cos t)$ remains bounded.
Now, let us consider \eqref{eq:p3}.
Even in the case when all the eigenvalues of $S$ have positive real parts, the initial condition
$$x_z(t_0) = \int_{t_0}^\infty e^{-S(t-t_0)} p x_1(t)dt$$
is well-defined because $x_1(t)$ is bounded by the stability of the closed-loop system with the controller.
Then, it can be shown that the solution $x_z(t)$ for $t \ge t_0$ is bounded.
However, it is not possible to predict such a value $x_z(t_0)$ because the future trajectory of $x_1(t)$, as well as $S$ and $p$, cannot be known at time $t_0$.
Nevertheless, the set of these particular initial conditions has measure zero, and so, the bounded solution of $x_z(t)$ is not likely to happen in practice.
\end{svgraybox}

Now, let us turn to the question how to estimate $a_a^*$.
In fact, \eqref{eq:realplant} can be written as follows with $a_a^*$:
\begin{align}\label{eq:my3}
\begin{split}
\dot x_: &= \ourA x_: + \ourB (\phi_:^\top x_: + \phi_z^\top x_z + b(u+a_a)) \\
&= \ourA x_: + \ourB (\phi_{:,\nn}^\top x_: + \phi_{z,\nn}^\top z_\attack + b_\nn u + b(a_a-a_a^*)).
\end{split}
\end{align}
If $a_a^*$ is treated as an unknown disturbance and $a_a$ is the control input, then this is the very situation where the disturbance observer technique can estimate $a_a^*$ with arbitrarily precision.
More specifically, if all the initial conditions at $t=0$ belong to a compact set, and if the disturbance $a_a^*(t)$ is bounded, then, for any $\epsilon>0$, a disturbance observer and an input $a_a$ can be designed such that 
\begin{equation}\label{eq:eps}
\|a_a(t) - a_a^*(t)\| \le \epsilon, \qquad \forall t \ge \epsilon.
\end{equation}
But, the signal $a_a^*(t)$ is not bounded if $x_z(t)$ is unbounded, which is the goal of attacker.
Therefore, precise estimation of $a_a^*$ by $a_a$ is not possible for infinite time horizon.
However, this is enough for attackers because, until $x_z(t)$ diverges up to the level of hazard in \eqref{eq:disruptive} at time $t^*$ (which takes a finite time duration $t^*-t_0$), the signal $a_a^*(t)$ can be treated as a bounded one and the disturbance observer can be designed to estimate it sufficiently closely.

Finally, the following dynamics generates robust zero-dynamics attack signal $a_a$, which approximates $a_a^*$ of \eqref{eq:rattack}, for $t_0 \le t \le t^*$:
\begin{align}\label{eq:robustattack}
    \begin{split}
        \dot z_\attack &= S_\nn z_\attack + p_\nn y \\
        \dot \xi_i &= \xi_{i+1} - \frac{q_{r-i}}{\tau^{i}} \xi_1 + \frac{q_0}{\tau^{r}} \frac{1}{b_\nn} \left( \phi_{:,\nn,r-i+1} + \frac{q_{r-i}}{\tau^i} \right) y, \qquad i = 1, \cdots, r-1, \\
        \dot \xi_r &= -\frac{q_0}{\tau^r} \xi_1 + \frac{q_0}{\tau^r} \frac{1}{b_\nn} \left( \phi_{:,\nn,1} + \frac{q_0}{\tau^r} \right) y + \frac{q_0}{\tau^r} \left( \frac{1}{b_\nn} \phi_{z,\nn}^\top z_\attack + u + a_a \right) \\
        a_a &= \sat_L \left( \xi_1 - \frac{q_0}{\tau^r}\frac{1}{b_\nn} y \right).
    \end{split}
\end{align}
All initial conditions at $t_0$ are set to zero, assuming that the plant is in the steady-state at time $t_0$.\footnote{As seen in \eqref{eq:p}, the effect of the proposed attack is to replace the real zero dynamics \eqref{eq:p3} with \eqref{eq:p2} at time $t_0$.
This is an abrupt change, and so, unless $z_\attack(t_0)$ is close to $x_z(t_0)$, some transient response may occur after $t_0$ and the attack may be detected.
To avoid this possibility, the attacker will carefully choose the time $t_0$ such that the internal state $x_z(t_0)$ is easily guessed, like the steady-state at which $z_\attack(t_0)=0$ may suffice.}
The function $\sat_L$ saturates at $\pm L$, and the reason for introducing saturation is that, even if the estimation of \eqref{eq:eps} holds for $t \ge t_0 + \epsilon$, nothing can be said for $t_0 \le t < t_0+\epsilon$.
In order to prevent unnecessary peak of $a_a(t)$ during this period (this peak tends to occur due to the peaking phenomenon \cite{sussmann1991peaking}, and by the peak, the trajectory \eqref{eq:my3} deviates from the attack-free trajectory hindering the stealthiness), we introduce the saturation function.
Now, the contants $q_i$ ($i=0,\cdots,r-1$), the level of saturation $L$, and the constant $\tau$ are design parameters, which are chosen from all the information available to the attacker.

\begin{proposition}
Let the polynomial $s^{r-1} + q_{r-1}s^{r-2} + \cdots + q_1$ be Hurwitz.
If $q_0>0$ is chosen sufficiently small, $L$ sufficiently large, and $\tau>0$ sufficiently small, then the actuator attack generated through \eqref{eq:robustattack} is stealthy.
If, in addition, the matrix $S$ contains at least one eigenvalue having positive real part, then the attack is disruptive for almost all initial conditions $x_z(t_0)$.
Finally, accuracy of estimation of $a_a^*$ by $a_a$ improves as $\tau$ gets smaller.
\end{proposition}

The proof is omitted due to space limit, but can be found in \cite{park_stealthy_2019}, where the details of choosing $q_0$, $L$, and $\tau$ are presented.
For more details on the disturbance observer technique used for \eqref{eq:robustattack}, refer to \cite{back2014reduced}.

\begin{example}

\begin{figure}[t]
\sidecaption
\includegraphics[width = 0.6\textwidth]{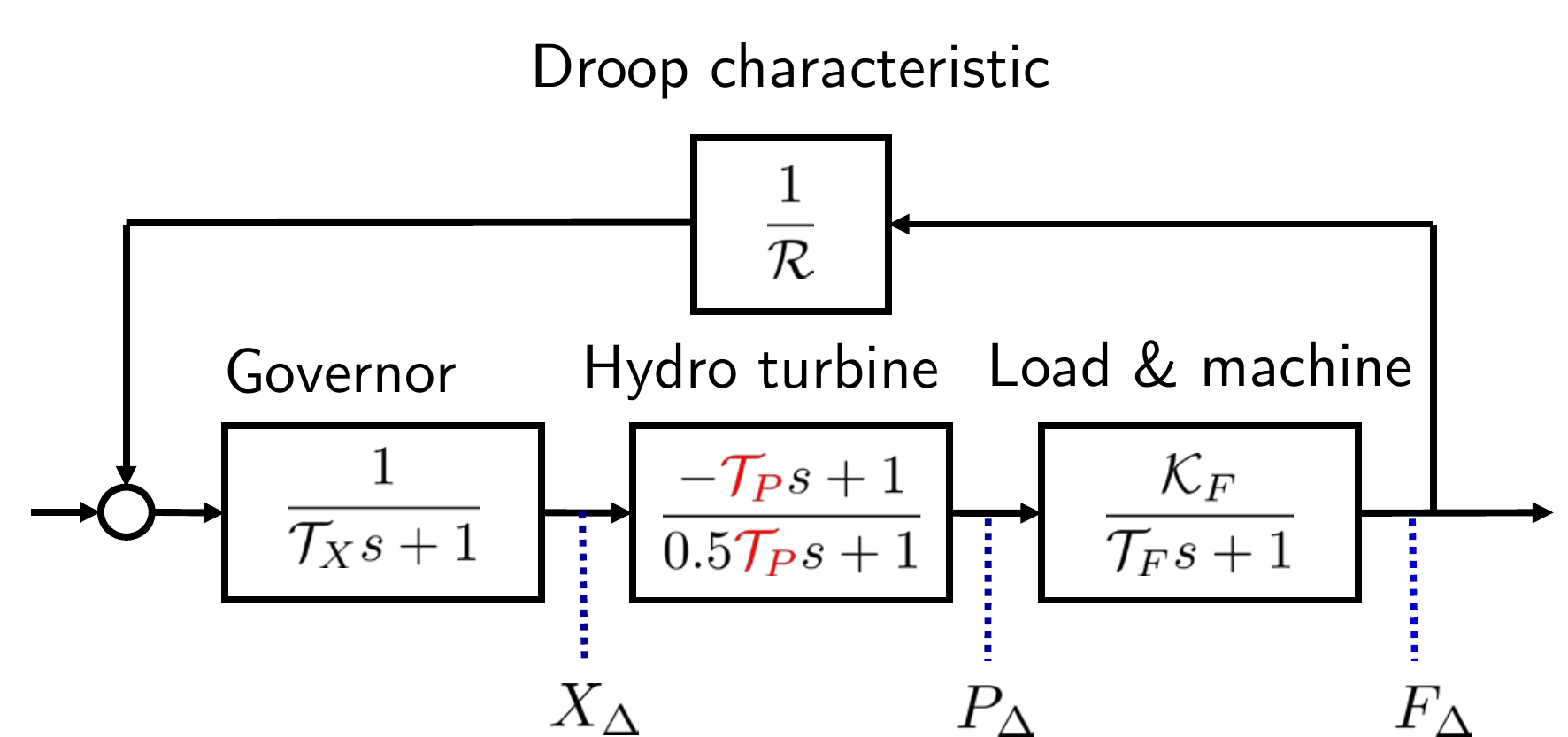}
\caption{Power generating plant with hydro turbine. 
$F_\Delta$: frequency deviation, $P_\Delta$: deviation in generated power, $X_\Delta$: deviation in governor's valve. 
The parameter $\TT_P$ is uncertain and belongs to the interval $[4,6]$.
Nominal $\TT_{P,\nn}$ is taken as $4$, and $\TT_X=0.2$, $\TT_F=6$, $\KK_F=1$, $\RR=0.05$.
(©2019 IEEE. Reprinted, with permission, from \cite{park_stealthy_2019})}
\label{fig:v4}
\end{figure}

A non-minimum phase uncertain system depicted in Fig.~\ref{fig:v4} is considered.
It is supposed that the plant is controlled by a PID-type controller $C(s) = (1.81s^2-18.9s+0.15)/(0.01s^2+s)$.
The classical zero-dynamics attack in Section \ref{sec:a} with the nominal parameter $\TT_{P,\nn}=4$ results in Fig.~\ref{fig:v5}.
The robust zero-dynamics attack in this section yields Fig.~\ref{fig:v6}.
For simulation, $q_0=1$, $L=20,000$, and $\tau=0.001$ are used.
\qed

\begin{figure}[t]
\sidecaption
\includegraphics[width = 0.5\textwidth]{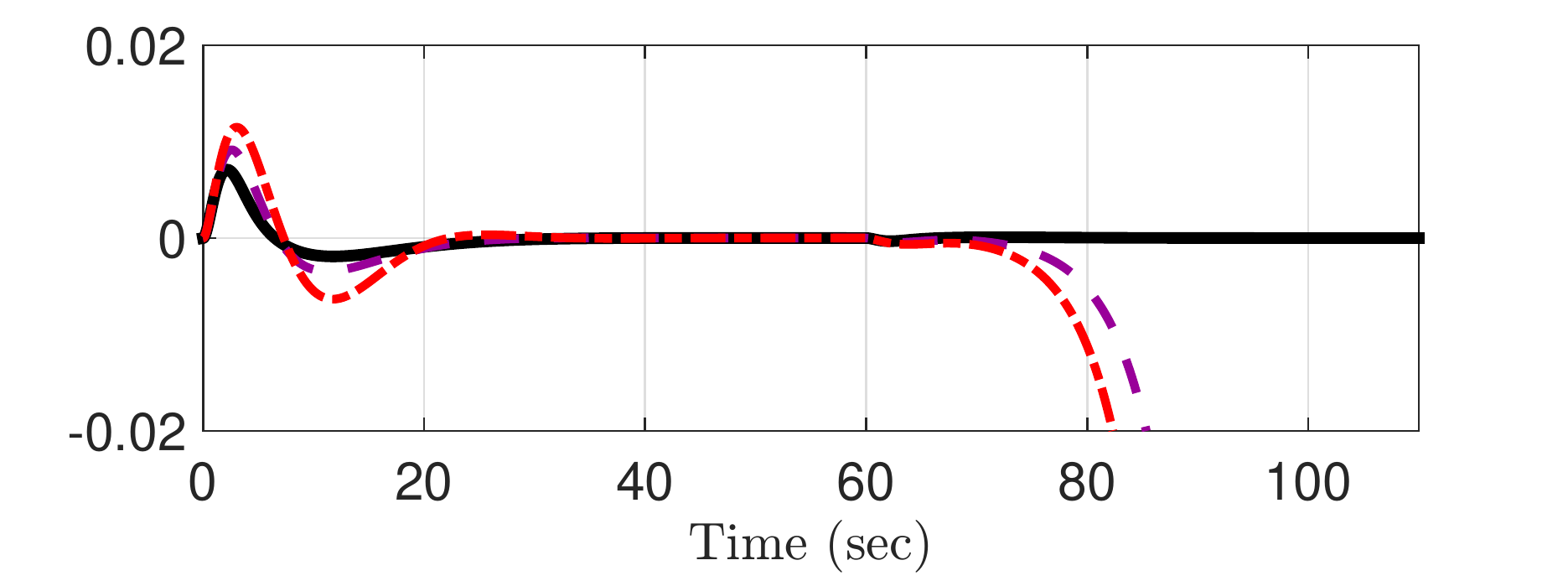} 
\includegraphics[width = 0.5\textwidth]{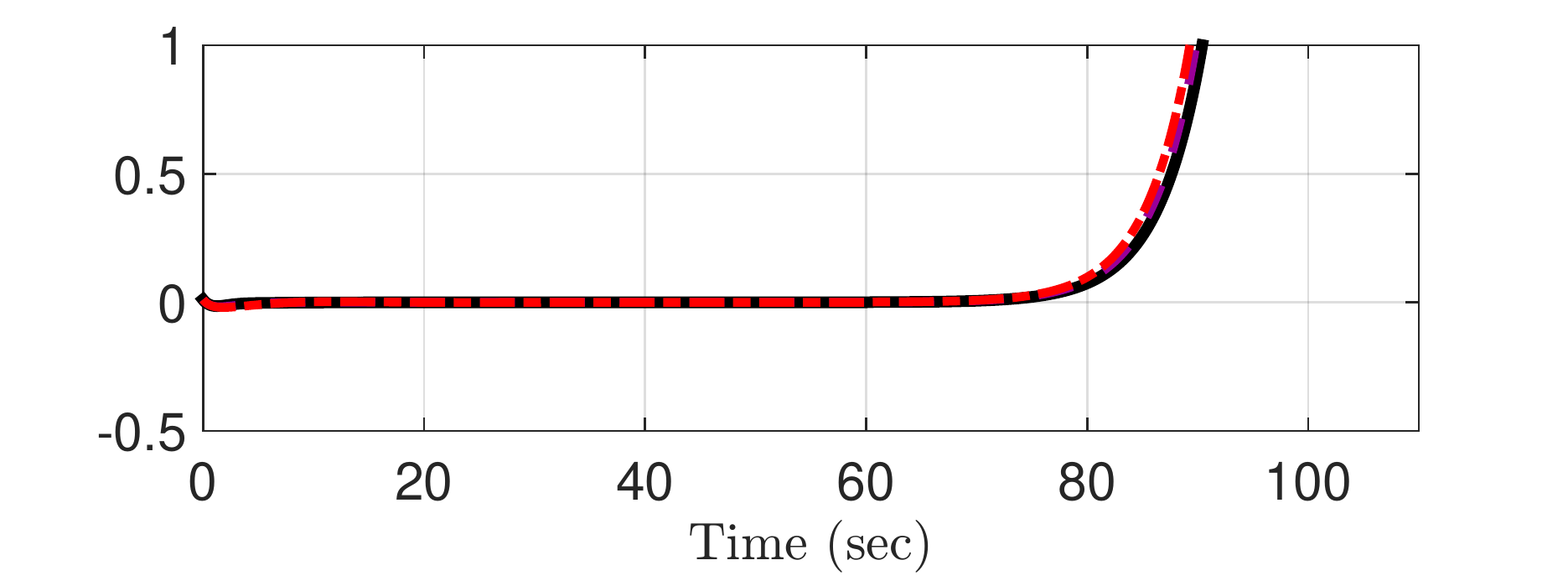}
\caption{Plots of the output $F_\Delta$ (left) and the state $X_\Delta$ (right).
Attack starts at 60 sec.
The black solid is the case when $\TT_P = \TT_{P,\nn}=4$ while the purple dash means $\TT_P=5$ and the red dash-dot means $\TT_P=6$.
With model uncertainty, the attack is easily detectable.
(©2016 IEEE. Reprinted, with permission, from \cite{park_when_2016})}
\label{fig:v5}
\end{figure}

\begin{figure}[t]
\includegraphics[width = 0.5\textwidth]{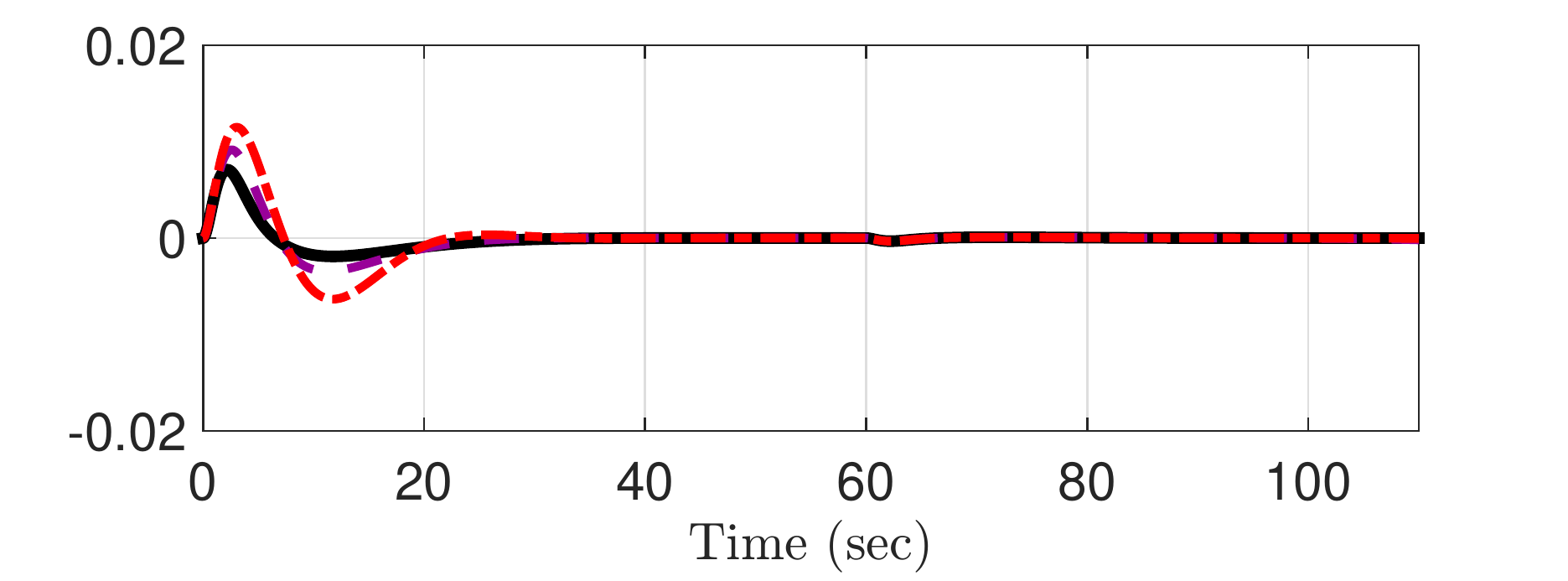}
\includegraphics[width = 0.5\textwidth]{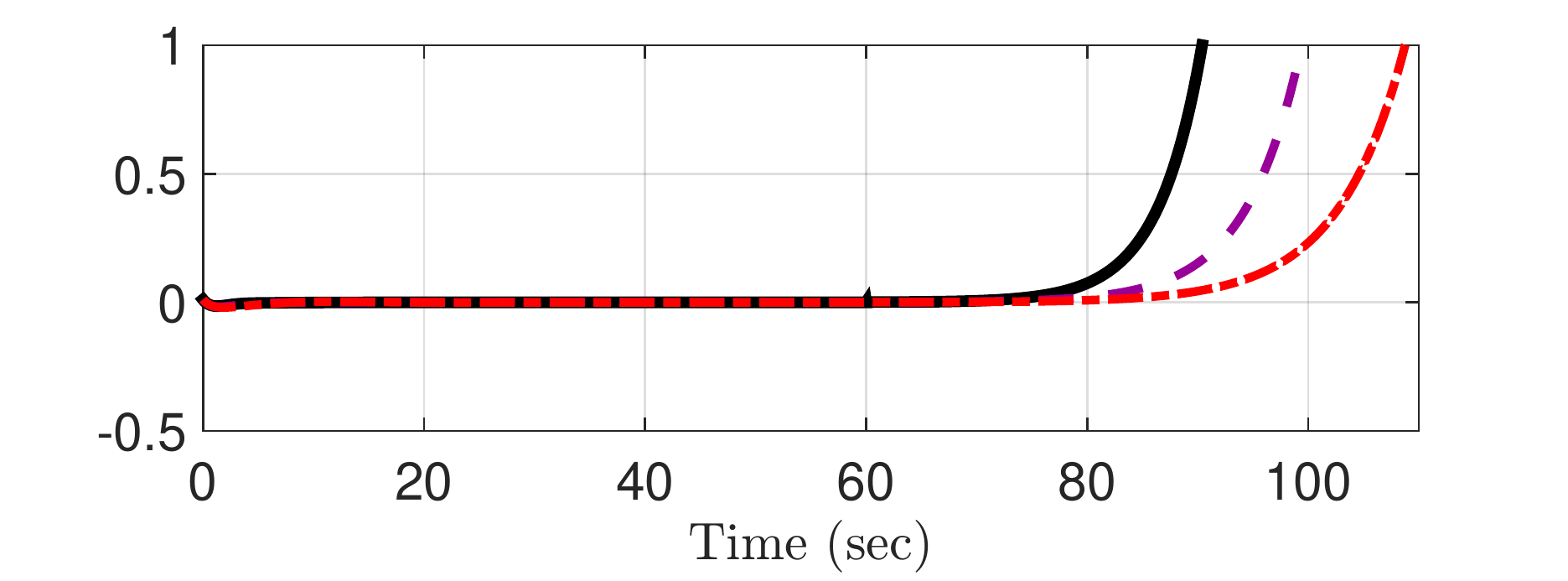}
\caption{Plots of the output $F_\Delta$ (left) and the state $X_\Delta$ (right).
Attack starts at 60 sec.
The black solid is the case when $\TT_P = \TT_{P,\nn}=4$ while the purple dash means $\TT_P=5$ and the red dash-dot means $\TT_P=6$.
In spite of model uncertainty, the attack remains stealthy.
(©2016 IEEE. Reprinted, with permission, from \cite{park_when_2016})}
\label{fig:v6}
\end{figure}

\end{example}

\subsubsection{Robust Pole-dynamics Attack}

As a dual to robust zero-dynamics attack, robust pole-dynamics attack is also possible.
Recall the closed-loop system in \eqref{eq:controller1}, which is written again as
\begin{align}\label{eq:pda1}
\begin{split}
    u = c_1 \qquad \dot c_: &= \ourA c_: + \ourB ( \phi_:^\top c_: + \phi_z^\top c_z + b (Cx + a_s) ) \\
    \dot c_z &= S c_z + p c_1 \\
    y = Cx \qquad \dot x &= A x + B c_1.
\end{split}
\end{align}
When the plant's parameters $A$, $B$, and $C$ have uncertainty and/or the controller's parameters $\phi_:$, $\phi_z$, $b$, $S$, and $p$ are uncertain, the robust pole-dynamics attack can be employed.
In particular, the zero dynamics part in \eqref{eq:pda1} and the uncertain parameters can be replaced with their nominal counterparts as
\begin{align}\label{eq:pda2}
\begin{split}
    \dot c_: &= \ourA c_: + \ourB ( \phi_{:,\nn}^\top c_: + \phi_{z,\nn}^\top z_{\attack,c} + b_\nn C_\nn z_{\attack,x} + b(a_s - a_s^*)) \\
    \begin{bmatrix} \dot z_{\attack,c} \\ \dot z_{\attack,x} \end{bmatrix} &= \begin{bmatrix} S_\nn & 0 \\ 0 & A_\nn \end{bmatrix} \begin{bmatrix} z_{\attack,c} \\ z_{\attack,x} \end{bmatrix} + \begin{bmatrix} p_\nn \\ B_\nn \end{bmatrix} c_1 
\end{split}
\end{align}
where $a_s^*$ can be estimated by $a_s$ through the disturbance observer as before.
For the details, refer to \cite{jeon_resilient_2016}.

\medskip

Up to now, only linear systems are considered as target systems. 
Is it possible to generate a disruptive stealthy attack for nonlinear systems?

\subsection{Zero-dynamics Attack for Nonlinear Systems}

In fact, nonlinear systems are also vulnerable to disruptive stealthy attacks.
When a single-input-single-output smooth nonlinear system has a well-defined relative degree $r$, it can be converted into the nonlinear normal form as
\begin{align}\label{eq:nonsys}
    \begin{split}
    y = x_1 \qquad 
    \dot x_: &= \ourA x_: + \ourB (f(x_:,x_z) + g(x_:,x_z) (u + a_a)) \\
    \dot x_z &= h(x_z,x_:)
    \end{split}
\end{align}
where $x_: \in \R^r$ and $x_z \in \R^{n-r}$.
For this system, while it is not straightforward to apply the approach of Section \ref{sec:a}, the approach of robust zero-dynamics attack can still be applied as long as $g(x_:,x_z)$ is a constant function such that $g(x_:,x_z)=b$, and $h(x_z,x_:)$ depends only on $(x_z,x_1)$.
With 
\begin{align*}
\dot z_\attack &= h_\nn(z_\attack,y) \\
a_a^* &= \frac{1}{b} \left( -f(x_:,x_z) - bu + f_\nn(x_:,z_\attack) + b_\nn u \right)    
\end{align*}
the system is rewritten as
\begin{align}
    \begin{split}
    y = x_1 \qquad 
    \dot x_: &= \ourA x_: + \ourB (f_\nn(x_:,z_\attack) + b_\nn u + b (a_a-a_a^*) ) \\
    \dot z_\attack &= h_\nn(z_\attack,x_1) \\
    \dot x_z &= h(x_z,x_1)
    \end{split}
\end{align}
Then, $a_a^*$ can be estimated by a nonlinear disturbance observer studied in \cite{back2008adding}.

\section{Disarming Zero-dynamics Attack}

In this section, we discuss how to protect the control system from the zero-dynamics attacks when it is implemented in the sampled-data framework.
In particular, we consider a SISO LTI continuous-time system
\begin{equation}\label{eq:CT_LTI}
\dot x(t) = Ax(t) + Bu(t), \qquad y(t) = Cx(t) 
\end{equation}
that is implemented in the sampled-data framework with 
\begin{equation}\label{eq:sampling}
u(t) = u[k] \quad \forall t \in [kT_s,(k+1)T_s), \qquad y[k] = y(kT_s).
\end{equation}
Because the attack cannot be detected by monitoring the sampled output $y[k]$, it is generally hard to conceive a countermeasure. 
Only a few protecting strategies are available in the literature. 
The idea of changing sensors and actuators is proposed in \cite{teixeira_revealing_2012}, and the idea of introducing a modulation block before the actuator in the feedback configuration is proposed in \cite{HoehnZhang2016ACC}. 
Both methods are intended to modify the system structure under the premise that the changes are kept confidential to the adversaries. 
Therefore, the security is compromised if the changes are disclosed to adversaries, and even if this is not the case, the sampling-zero-dynamics attack may still be effective when the sampling period $T_s$ is very small.
This is because, if $T_s$ is very small, the sampling zeros approach the roots of Euler-Frobenius polynomial, and the Euler-Frobenius polynomial (which is known to the public) does not depend on the particular parameters of the target system (as long as the relative degree of the system is known).

In this section, we present another strategy that can disarm the zero-dynamics attack.
The idea is to replace the conventional zero-order-hold or the sampler in the plant-side with a generalized hold (GH) or a generalized sampler (GS), respectively.\footnote{The generalized hold and sampler have been actively studied in the 90s. For the details, refer to \cite{YuzGoodwin2014Book}.}
With this change, it is intended that the unstable zeros are relocated to the stable region in the discrete-time domain.
Therefore, if the change is not known to the adversaries then the attack will be detected, and even if the change is known so that the attack is made stealthy, the attack is no more disruptive because all the zeros are stable.
Eventually, the adversaries lose the motivation of attack.

\subsection{Moving Zeros by Generalized Hold}\label{sec:3.1}

In this subsection, we introduce the idea of using generalized hold (GH), which is based on the work of \cite{kim_neutralizing_2020}.
Generalized hold \cite{YuzGoodwin2014Book} is a device that generates the continuous-time input to the plant from a given input sequence $u[k]$ as 
\begin{equation}\label{eq:Proto_GH}
u_g(t) = h_g(t-kT_s) u[k] \qquad \forall t \in [kT_s,(k+1)T_s)
\end{equation}
where $h_g$ is a function defined on $[0,T_s)$.
Recall that if $h_g(t)=1$ for $[0, T_s)$, then GH recovers zero-order-hold.  
With GH, the sampled-data model of \eqref{eq:CT_LTI} becomes
\begin{equation}\label{eq:DT_GH}
x[k+1] = A_{\mathsf d} x[k] + B_g u[k], \quad y[k] = C_{\mathsf d} x[k]
\end{equation}
where $x[k]\in \R^n$ and
$$A_{\mathsf d}= e^{AT_s}, \quad B_g := \int_0^{T_s} e^{A(T_s-\tau)}Bh_g(\tau) d\tau, \quad C_{\mathsf d}=C.$$ 
The sampled-data transfer function from $u[k]$ to $y[k]$, denoted by $G_{\mathsf d}(z)$, is then given by 
$$G_{\mathsf d}(z)= C_{\mathsf d} (zI_n - A_{\mathsf d} )^{-1}B_g.$$
Then, in most cases, by a suitable design of $h_g$, the zeros of $G_{\mathsf d}(z)$ can be assigned to arbitrary locations.
To see this, let 
\begin{equation}\label{eq:G^*}
G^*_{\mathsf d}(z)=k_{\mathsf d}\frac{(z-z_{\mathsf d,1})\cdots(z-z_{\mathsf d,n-1})}{{\rm det}(zI_n- A_{\mathsf d})}
\end{equation}
be a desired transfer function where $z_{\mathsf d,1}, \dots, z_{{\mathsf d},n-1} \in \mathbb C$ are the desired zeros (in complex conjugate pairs) and $k_{\mathsf d}$ is a desired gain. 

\begin{lemma}\label{lem:B_g}
Suppose $(A_{\mathsf d}, C_{\mathsf d})$ is observable.\footnote{It is well-known that, if $(A,C)$ is observable, then $(A_{\mathsf d}, C_{\mathsf d})$ is observable for almost all sampling times $T_s$.} Then, there exists a unique $B_g\in \R^n$ such that $G_{\mathsf d}(z) = G^*_{\mathsf d}(z)$. 
\end{lemma}

\begin{proof}
Let $C_* (z I_n - A_*)^{-1} B_*$ is a minimal realization of $G_{\mathsf d}^*(z)$.
Then, $G_{\mathsf d}(z)$ is identical to $G^*_{\mathsf d}(z)$ if and only if $C_{\mathsf d} A_{\mathsf d}^{k-1}B_g = C_* A_*^{k-1} B_*$ for all $k= 1,2,\dots,n$.
With ${\mathcal O}_{\mathsf d}$ being the observability matrix of $(A_{\mathsf d}, C_{\mathsf{d}})$, it is equivalent to
\begin{equation}\label{eq:Bg}
{\mathcal O}_{\mathsf d} B_g = \begin{bmatrix} C_\ddd \\ C_\ddd A_\ddd \\ \vdots \\ C_\ddd A_\ddd^{n-1} \end{bmatrix} B_g = \begin{bmatrix} C_* B_* \\ C_* A_* B_* \\ \vdots \\ C_* A_*^{n-1} B_* \end{bmatrix}.
\end{equation}
By the observability, ${\mathcal O}_{\mathsf d}$ is invertible.
\qed
\end{proof}

With $B_g$ obtained above, it remains to construct $h_g(t)$ such that 
\begin{equation}\label{eq:Bghg}
B_g=\int_0^{T_s}e^{A(T_s - \tau)} B h_g(\tau)d\tau.
\end{equation}
Then, it is immediate to see that such an $h_g$ exists if $(A,B)$ is controllable because the form \eqref{eq:Bghg} reminds, for $\dot x = A x + B u$, the problem of driving the state from the origin to the vector $B_g$ at time $T_s$ by the input $u(t) = h_g(t)$.
So a solution is given by
\begin{equation}\label{eq:h_g_cont}
h_g(t) = B^\top e^{A^{\top}(T_s-t)} W^{-1}(0,T_s)B_g
\end{equation}
where $W(0,T_s)$ is the controllability Gramian $W(0,T_s)=\int_{0}^{T_s} e^{A\tau} BB^\top e^{A^\top \tau} d\tau$.
However, this requires a device that can generates continuous-time signals, which may not be very practical.

Alternatively, a piecewise constant function $h_g$ becomes of interest.
Let $N$ be the number of subintervals within which $h_g(t)$ is constant, namely
\begin{equation}\label{eq:PCGH}
h_g(t) = h_i, \qquad \frac{(i-1)T_s}{N} \le t < \frac{iT_s}{N}, \quad i = 1,\dots,N .
\end{equation}
Substituting \eqref{eq:PCGH} into \eqref{eq:Bghg}, one has 
\begin{equation*}\label{eq:Bghg_PC}
B_g = \sum_{i=1}^N \int_{\frac{(i-1)T_s}{N}}^\frac{iT_s}{N}e^{A(T_s-\tau)}B d\tau \; h_i,
\end{equation*}
which can be simply written, with $\bar h := [h_1, \cdots, h_N ]^\top$, as
\begin{equation}\label{eq:Ch}
B_g = \begin{bmatrix} A_{\mathsf{d},N}^{N-1} B_{\mathsf{d},N}, & \cdots, & A_{\mathsf{d},N} B_{\mathsf{d},N}, & B_{\mathsf{d},N} \end{bmatrix} \bar h =: M_B \bar h
\end{equation}
where $A_{\mathsf{d},N} = e^{A T_s/N}$, $B_{\mathsf{d},N} = \int_0^{T_s/N} e^{A(T_s/N -\tau)} B d\tau$, and $M_B \in \mathbb{R}^{n\times N}$.

Therefore, we obtain the following result.
\begin{lemma}\label{lem:ZeroAssignPCGH}
If $B_g$ belongs to the range space of $M_B$, then the vector $\bar h$ exists.
If $(A_{\mathsf{d},N},B_{\mathsf{d},N})$ is controllable (which is the case for almost all $T_s$ when $(A,B)$ is controllable), then the vector $\bar h$ exists for any $B_g$ with $N=n$, and $\bar h = M_B^{-1} B_g$.
\end{lemma}

\subsection{Optimal Design of Generalized Hold}\label{sec:DesignGH}

The approach of moving the system zeros to the stable region by GH is effective in disarming the zero-dynamics attack. 
However, the response of the resulting system with GH is different from that with ZOH. 
In other words, the closed-loop system performance is affected by designing the GH and it may degrade from the original design. 
The question  ``how can we design GH so that the performance degradation of the closed-loop system (from that with ZOH) is minimized?"  naturally follows, which is the topic of this subsection.
  
In order to design GH following the procedure developed in the previous subsection, one needs to choose the desired zeros first, and computes $B_g$ and then $h_g$ in order.
But, then, the values of $h_g(t)$ may happen to be very different from those of the ZOH (which are $1$).
To overcome this drawback, by recalling that our interest is not to assign the zeros at specific locations but to make the system have stable discrete-time zeros, we formulate an optimization problem to minimize the difference between $h_g(t)$ and $1$, under the constraints that the zeros are located inside the unit circle and that the integrations of $h_g(t)$ and 1 during one sampling period are equal.

For building the stability constraint, we note that the realization of $G_{\mathsf d}^*(z)$ of \eqref{eq:G^*} in the controllable canonical form is given by
\begin{align*}
{\bar x}[k+1] &= \left[\begin{array}{c} 
\begin{tabular}{c|c}$0_{n-1}$& $I_{n-1}$\end{tabular}  \\ \hline \begin{matrix} -d_{0} & \cdots & -d_{n-1} \end{matrix}\end{array} \right]{\bar x}[k] 
+ \begin{bmatrix} 0_{n-1}\\ 1\end{bmatrix}u[k]  =: A_* {\bar x}[k]+ B_* u[k]\\
y[k] &= \begin{bmatrix} c_0 &\cdots& c_{n-2}& c_{n-1} \end{bmatrix}\bar x[k] =: C_* \bar x[k]
\end{align*}
where the constants $c_{0}$, $\dots$, $c_{n-1}$, $d_{0}$, $\dots$, $d_{n-1}$ are determined from the relations 
\begin{align}
{\rm det}(zI_n -A_{\mathsf d} )&= z^n + d_{n-1} z^{n-1} +\cdots+ d_{0} \notag \\  
k_{\mathsf d}\prod_{i=1}^{n-1}(z-z_{\mathsf d, i}) &= c_{n-1}z^{n-1} + c_{n-2} z^{n-2} + \cdots + c_0. \label{eq:stablepoly}
\end{align}
Then, it follows from \eqref{eq:Bg} and \eqref{eq:Ch} that
\begin{equation}\label{eq:commute}
{\mathcal O}_{\mathsf d} M_B \bar h = {\mathcal O}_{\mathsf d} B_g = \begin{bmatrix} C_* B_* \\ \vdots \\ C_* A_*^{n-1} B_* \end{bmatrix} 
= \begin{bmatrix} B_*^\top \\ \vdots \\ B_*^\top (A_*^\top)^{n-1} \end{bmatrix} C_*^\top =: {\mathcal C}_*^\top C_*^\top
\end{equation}
where ${\mathcal C}_*$ is the controllability matrix of $(A_*,B_*)$, and we used the fact that $C_* A_*^{i-1} B_*$ is scalar in the derivation.

\begin{remark}\label{rem:ht}
Suppose that we seek for a Schur stable polynomial of order $q$; i.e., $\alpha_q z^q + \alpha_{q-1} z^{q-1} + \cdots + \alpha_0$.
By utilizing the LMI condition for a strict positive real system, an LMI specification for the search can be formulated by
\begin{align*}
&\text{find $\tilde C = [\alpha_0, \alpha_1, \cdots, \alpha_{q-1}]$, $\tilde D = \alpha_q > 0$ and $\tilde P>0$} \\
&\text{such that} \quad \begin{bmatrix} \tilde A^\top \tilde P \tilde A - \tilde P, & \tilde A^\top \tilde P \tilde B - \tilde C^\top \\ \tilde B^\top \tilde P \tilde A - \tilde C, & \tilde B^\top \tilde P \tilde B - 2\tilde D \end{bmatrix} \le 0,
\end{align*}
where 
$$\tilde A = \begin{bmatrix} 0 & 1 & 0 & \cdots & 0 \\ 0 & 0 & 1 & \cdots & 0 \\
\vdots & \vdots & \vdots & \ddots & \vdots & \\
0 & 0 & 0 & \cdots & 1 \\ 0 & 0 & 0 & \cdots & 0 \end{bmatrix} \in \R^{q \times q}, \qquad
\tilde B = \begin{bmatrix} 0 \\ 0 \\ \vdots \\ 0 \\ 1 \end{bmatrix} \in \R^q.$$
The underlying idea is to ask the solver to search for $\tilde C$ and $\tilde D$ such that
$$\tilde C (zI-\tilde A)^{-1} \tilde B + \tilde D = \frac{\alpha_q z^q + \cdots + \alpha_0}{z^q}$$
becomes strictly positive real (which is already stable since $\tilde A$ is Schur) \cite{Hitz}.
\qed
\end{remark}

Based on the idea of Remark \ref{rem:ht}, an optimization problem is formulated as
\begin{subequations}\label{eq:opt_prob0}
\begin{align}
&\displaystyle{\minimize_{\bar h \in \R^N, \; \tilde P \in \R^{(n-1)\times(n-1)}}} \;\; \|\bar h - 1_N\|^2 \\
&\qquad \text{subject to} \;\; \begin{bmatrix} \tilde A^\top \tilde P \tilde A - \tilde P, & \tilde A^\top \tilde P \tilde B \\ \tilde B^\top \tilde P \tilde A, & \tilde B^\top \tilde P \tilde B \end{bmatrix} - ({\mathcal C}_*^{-\top} {\mathcal O}_{\mathsf d} M_B \bar h) e_n^\top - e_n ({\mathcal C}_*^{-\top} {\mathcal O}_{\mathsf d} M_B \bar h)^\top \le 0 \label{eq:h2} \\
&\qquad\quad\quad \text{and} \;\; \tilde P > 0, \label{eq:h3} \\
&\qquad\quad\quad \text{and} \;\; 1_N^\top \bar h = N, \label{eq:h4}
\end{align}
\end{subequations}
where $1_N \in \R^N$ is the vector with all elements being 1, and $e_n \in \R^n$ is the elementary vector whose $n$-th element is 1 while all others are zero.
Indeed, the column vector $({\mathcal C}_*^{-\top} {\mathcal O}_{\mathsf d} M_B \bar h)$ is nothing but $[c_0, c_1, \cdots, c_{n-1}]^\top$ due to \eqref{eq:stablepoly} and \eqref{eq:commute}.
The task of disarming the attack is achieved mainly by \eqref{eq:h2} and \eqref{eq:h3}, and thus, the cost function and the constraint \eqref{eq:h4} can be modified to obtain different $\bar h$.

\subsection{Moving Zeros by Generalized Sampler}

In this subsection, it is shown that, by using a generalized sampler (GS), the zeros of the discrete-time representation of the plant can also be placed inside the unit circle. 
Zero-dynamics attack may remain stealthy, but since all attack signals diminish, it becomes not disruptive. 
Along with the approach of using GH, the use of GS disarms the zero-dynamics attack. 
This subsection is based on the work of \cite{Kim+2020IFACWC}.

Generalized sampler \cite{YuzGoodwin2014Book} is a device that generates the output sample from the continuous-time output of the plant as
\begin{equation}\label{eq:Proto_GS}
y_g[k] = \int_0^{T_s} h_g(t) y(t+(k-1)T_s) dt
\end{equation}
where $h_g$ is a (generalized) function defined on $(0,T_s]$.
The conventional sampler $y[k] = y(kT_s)$ can be considered as a GS with $h_g(t) = \delta(t-T_s)$.
Recalling that\footnote{Note that $\int_0^t e^{A(t-\tau)} B d\tau = \int_0^t e^{A\tau} B d\tau$.} $y(t+(k-1)T_s) = Ce^{At}x[k-1] + C \int_0^t e^{A\tau} B d\tau \; u[k-1]$ for $0 < t \le T_s$, it is seen that
\begin{align}\label{eq:compare}
\begin{split}
y_g[k] &= \left( \int_0^{T_s} h_g(t) C e^{At} dt \right) x[k-1] + \left( \int_0^{T_s} h_g(t) C \int_0^t e^{A\tau}B d\tau dt \right) u[k-1] \\
&=: C_g \; x[k-1] + D_g \; u[k-1].
\end{split}
\end{align}
With $x[k+1] = e^{AT_s} x[k] + \int_0^{T_s} e^{A\tau} B d\tau u[k] =: A_{\mathsf d} x[k] + B_{\mathsf d} u[k]$,
the transfer function from $u[k]$ to $y_g[k+1]$ is given by $C_g (zI_n-A_{\mathsf d})^{-1} B_{\mathsf{d}} + D_g$, and therefore, the transfer function from $u[k]$ to $y_g[k]$ becomes
\begin{align}\label{eq:NewSDtf}
G_{\mathsf{d}}(z) = z^{-1}(C_g (zI_n-A_{\mathsf d})^{-1} B_{\mathsf{d}} + D_g)
\end{align}
which has dimension $n+1$ (rather than $n$) due to the one-step delay.
Then, in most cases, by a suitable design of $h_g$, the zeros of $G_{\mathsf d}(z)$ can be assigned to arbitrary locations.
To see this, let 
\begin{align}\label{eq:desiredtf}
G^*_{\mathsf{d}}(z) = k_{\mathsf d} z^{-1} \frac{(z - z_{\mathsf d,1}) \cdots (z-z_{\mathsf d,n})}{{\rm det}(zI_n- A_{\mathsf{d}})}
\end{align}
be a desired transfer function where $z_{\mathsf d,1}, \dots, z_{{\mathsf d},n} \in \mathbb C$ are the desired zeros (in complex conjugate pairs) and $k_{\mathsf d}$ is a desired gain. 

\begin{lemma}\label{lem:C_g}
Suppose $(A_{\mathsf d}, B_{\mathsf d})$ is controllable.\footnote{It is well-known that, if $(A,B)$ is controllable, then $(A_{\mathsf d}, B_{\mathsf d})$ is controllable for almost all sampling times $T_s$.} Then, there exists a unique pair $C_g \in \R^{1 \times n}$ and $D_g \in \R$ such that $G_{\mathsf d}(z) = G^*_{\mathsf d}(z)$. 
\end{lemma}

\begin{proof}
Let $C_* (z I_n - A_*)^{-1} B_* + D_*$ is a minimal realization of $z G_{\mathsf d}^*(z)$.
Then, $z G_{\mathsf d}(z)$ is identical to $z G^*_{\mathsf d}(z)$ if and only if $D_g = D_*$ and $C_g A_{\mathsf d}^{k-1}B_{\mathsf d} = C_* A_*^{k-1} B_*$ for all $k= 1,2,\dots,n$.
With ${\mathcal C}_{\mathsf d}$ being the controllability matrix of $(A_{\mathsf d}, B_{\mathsf{d}})$, it is equivalent to
\begin{align}\label{eq:Cg}
\begin{split}
C_g {\mathcal C}_{\mathsf d} &= C_g \begin{bmatrix} B_\ddd, A_\ddd B_\ddd, \cdots, A_\ddd^{n-1} B_\ddd \end{bmatrix} \\
&= [C_* B_*, \; C_* A_* B_*, \; \cdots, \; C_* A_*^{n-1} B_* ] \qquad \text{and} \qquad D_g = D_* = k_{\mathsf d}.
\end{split}
\end{align}
By the controllability, ${\mathcal C}_{\mathsf d}$ is invertible.
\qed
\end{proof}

From now on, let us consider the case when $h_g(t) = \sum_{i=1}^N w_i \delta(t - iT_s/N)$, where $N$ is a positive integer and $w_i$'s are constant weights, so that
\begin{equation}\label{eq:newyk}
y_g[k] = \sum_{i=1}^N w_i y\left( \frac{i}{N}T_s + (k-1) T_s \right).
\end{equation}
While \eqref{eq:Proto_GS} requires continuous measurements over the interval, only finite number of measurements are asked in \eqref{eq:newyk}, which is practical.\footnote{A similar strategy was presented in \cite{Naghnaeian+2019SCL} where a multi-rate sampler is employed for attack detection. However, all samples $y(iT_s/N + (k-1)T_s)$, $i=1, \cdots, N$, are transmitted to the controller for attack detection.}

Now let us look at the relation between the pair $(C_g,D_g)$ and $\bar w := [w_1, \cdots, w_N]^\top$.
For this, note that, for $i=1,\cdots,N$,
\begin{align}\label{eq:hoo}
\begin{split}
x\left( \frac{i}{N} T_s + (k-1)T_s \right) &= e^{A\frac{iT_s}{N}} x[k-1] + \int^{\frac{iT_s}{N}}_0 e^{A(\frac{iT_s}{N} - \tau)} B d\tau \; u[k-1] \\
&= A_{\mathsf{d},N}^i \; x[k-1] + \sum_{j=1}^i A_{\mathsf{d},N}^{j-1} B_{\mathsf{d},N} \; u[k-1]
\end{split}
\end{align}
where $A_{\mathsf{d},N} = e^{A T_s/N}$ and $B_{\mathsf{d},N} = \int_0^{T_s/N} e^{A(T_s/N -\tau)} B d\tau = \int_0^{T_s/N} e^{A\tau} B d\tau$.
Then, by comparing \eqref{eq:newyk} and \eqref{eq:hoo} with \eqref{eq:compare}, it follows that
\begin{equation}\label{eq:compare2}
C_g = \bar w^\top \begin{bmatrix} C A_{\mathsf{d},N} \\ C A_{\mathsf{d},N}^2 \\ \vdots \\ C A_{\mathsf{d},N}^{N} \end{bmatrix} =: \bar w^\top M_C, \quad
D_g = \bar w^\top \begin{bmatrix} C B_{\mathsf{d},N} \\ C A_{\mathsf{d},N} B_{\mathsf{d},N} + C B_{\mathsf{d},N} \\ \vdots \\ C \sum_{j=1}^{N} A_{\mathsf{d},N}^{j-1} B_{\mathsf{d},N} \end{bmatrix} =: \bar w^\top M_D, 
\end{equation}
where $M_C \in \R^{N \times n}$ and $M_D \in \R^{N \times 1}$.
Therefore, we obtain the following result.

\begin{lemma}\label{lemma:existence}
If $[C_g,D_g] \in \R^{1 \times (n+1)}$ belongs to the row space of $[M_C,M_D] \in \R^{N \times (n+1)}$, then the vector $\bar w$ exists as a solution to
\begin{equation}\label{eq:MCD}
\bar w^\top [M_C, M_D] = [C_g, D_g].
\end{equation}
If the extended system
$$\tilde x[k+1] = \begin{bmatrix} A_{\mathsf{d},N} & B_{\mathsf{d},N} \\ 0 & 1 \end{bmatrix} \tilde x[k], \qquad \tilde y[k] = \begin{bmatrix} C A_{\mathsf{d},N}, & C B_{\mathsf{d},N} \end{bmatrix} \tilde x[k]$$
where $\tilde x \in \R^{n+1}$ is observable, then the vector $\bar w$ exists for any $[C_g,D_g]$ with $N = n+1$, and obtained by $\bar w^\top = [C_g,D_g] [M_C, M_D]^{-1}$.
\end{lemma}

\begin{proof}
The first claim is straightforward, and the second claim follows from the fact that the observability matrix of the extended system is $[M_C,M_D]$.
\qed
\end{proof}

\subsection{Optimal Design of Generalized Sampler}

Unlike the undesirable inter-sample behavior caused by the generalized hold, the generalized sampler does not affect the inter-sample behavior of the continuous-time system.
Instead, the output $y_g[k]$ is different from the conventional output sample $y[k]$, which may not be desirable for monitoring or for other purposes.
To mitigate this difference, an optimization problem can be formulated.\footnote{This subsection is a brief summary of the contribution of \cite{Kim_submit}.}
In particular, we want to reduce $\|y_g[k] - y[k]\|$ under the constraints that the zeros are located inside the unit circle and that the sum of the weights is unity, i.e., $1_N^\top \bar w = 1$.
Reducing the difference $\|y_g[k] - y[k]\|$ is related to the selection of $\bar w$ because
\begin{align*}
y_g[k] = \bar w^\top [M_C, M_D] \begin{bmatrix} x[k-1] \\ u[k-1] \end{bmatrix} \quad \text{and} \quad 
y[k] = e_N^\top [M_C, M_D] \begin{bmatrix} x[k-1] \\ u[k-1] \end{bmatrix}
\end{align*}
where $e_N = [0, \cdots, 0, 1]^\top \in \R^N$, in which the left equation follows from \eqref{eq:compare} and \eqref{eq:compare2}, and the right equation is from \eqref{eq:newyk} by observing that \eqref{eq:newyk} becomes $y[k] = y(kT_s)$ when $w_N = 1$ and $w_i = 0$ for all $i \not = N$.

Now, we need to relate $\bar w$ with the stability of the zero polynomial in \eqref{eq:desiredtf}.
This is achieved by realizing $z G_{\mathsf{d}}^*(z)$ in the controllable canonical form as
\begin{align*}
\bar x[k+1] &= \left[ \begin{array}{c} \begin{array}{c|c} 0_{n-1}\  &\  I_{n-1} \\ \end{array} \\
\hline 	-d_{0} \,\, \dotsb \,\, -d_{n-1}    \end{array}\right] \bar x[k] +
\begin{bmatrix}    0_{n-1}\\  1 \end{bmatrix} u[k] =: A_* \bar x[k] + B_* u[k] \\ 
y_g[k+1] &= \begin{bmatrix}c_{0}& \dotsb & c_{n-2} & c_{n-1}\end{bmatrix} \bar x[k] + k_{\mathsf{d}} u[k] =: C_* \bar x[k] + D_* u[k] 
\end{align*}
where $D_* = k_{\mathsf{d}}$ and the constants $c_{0}$, $\dots$, $c_{n-1}$, $d_{0}$, $\dots$, $d_{n-1}$ are determined from the relations 
\begin{align}\label{eq:stable3}
\begin{split}
{\rm det}(zI_n -A_{\mathsf d})&= z^n + d_{n-1} z^{n-1} +\cdots+ d_{0} \\  
k_{\mathsf{d}} \left( \prod_{i=1}^{n}(z-z_{\mathsf{d}, i}) -{\rm det}(zI_n -A_{\mathsf d}) \right) &= c_{n-1}z^{n-1} + c_{n-2} z^{n-2} + \cdots + c_0 .
\end{split}
\end{align}
With $C_* = [c_0, \cdots, c_{n-1}]$ and $D_* = k_\ddd$, it follows from \eqref{eq:MCD} and \eqref{eq:Cg} that 
\begin{align*} 
\bar w^\top M_C &= C_g = \left[ C_* B_*, \; \cdots, \; C_* A_*^{n-1} B_* \right] {\mathcal C}_{\mathsf d}^{-1} = C_* \left[ B_*, \; \cdots, \; A_*^{n-1} B_* \right] {\mathcal C}_{\mathsf d}^{-1} \\
&=: C_* {\mathcal C}_* {\mathcal C}_{\mathsf d}^{-1}, \\
\bar w^\top M_D &= D_g = D_* = k_{\mathsf{d}}.
\end{align*}

Finally, the convex optimization problem is given by
\begin{subequations}\label{eq:opt_GS}
\begin{align}
\displaystyle{\minimize_{\bar w \in \R^N, \; \tilde P \in \R^{n \times n}}} \;\; & \| [M_C,M_D]^\top (\bar w - e_N) \|^2 \\
\text{subject to} \;\; & 
\begin{bmatrix} \tilde A^\top \tilde P \tilde A - \tilde P, & \tilde A^\top \tilde P \tilde B - \tilde C^\top \\ \tilde B^\top \tilde P \tilde A - \tilde C, & \tilde B^\top \tilde P \tilde B - 2\tilde D \end{bmatrix} \le 0, \quad \tilde P > 0, \label{eq:w1} \\
\text{and} \;\; & \tilde C = {\mathcal C}_*^{-\top} {\mathcal C}_{\mathsf{d}}^\top M_C^\top \bar w + (M_D^\top \bar w) \bar d, \quad \tilde D = M_D^\top \bar w \label{eq:w2} \\
\text{and} \;\; & 1_N^\top \bar w = 1 \label{eq:w4}
\end{align}
\end{subequations}
where $\bar d := [d_0, d_1, \cdots, d_{n-1}]^\top$.
The constraints \eqref{eq:w1} and \eqref{eq:w2} are for stability of the zero polynomial $k_{\mathsf d} \prod_{i=1}^{n}(z-z_{\mathsf{d}, i})$, derived from \eqref{eq:stable3} and Remark \ref{rem:ht}.
The task of disarming the attack is achieved mainly by \eqref{eq:w1} and \eqref{eq:w2}, and thus, the cost function and the constraint \eqref{eq:w4} can be modified to obtain different $\bar w$.

\section{Conclusion}

This chapter introduces an actuator attack that is both stealthy and disruptive. In particular, the attack exploits the zero dynamics of the system, which is either from  intrinsic zeros or from sampling zeros. Also presented is the robust zero-dynamics attack that is both disruptive and stealthy but the design of which does not require exact knowledge of the target system. This implies that the target control system may be in a greater danger than one would imagine.  
The attack extends to the case of nonlinear systems. 
As a dual problem of the zero-dynamics attack, two sensor attacks are briefly introduced as well, which are referred to as pole-dynamics attack and robust pole-dynamics attack, respectively. These are attacks on sensors that exploit the plant pole dynamics.
Finally, two methods are introduced that render the zero-dynamics attack to be  little of a threat. It is shown that by designing a generalized hold or sample, unstable zeros move to a stable region in the complex plane. 
Thus, the attacks designed for the zeros are no longer disruptive. 
This kind of solutions are effective compared to other countermeasures against zero-dynamics attack, but require a redesign of feedback controllers because the discrete-time transfer function is changed by the generalized hold or sampler.
In addition, robustness of the designed hold or sampler against uncertainty of the plant needs to be investigated as the issue has been raised in, e.g., \cite{freud97}.

\section*{Acknowledgement}

The authors are grateful to Hyuntae Kim at Seoul National University for his idea of Remark \ref{rem:ht}.
This work was supported by Institute for Information \& communications Technology Promotion grant funded by MSIT, the Korean government (2014-0-00065, Resilient Cyber-Physical Systems Research).

\end{document}